\global\def\draftcontrol{0}

   \def\versionno{ ks bh }

\catcode`\@=11

\expandafter\ifx\csname draftcontrol\endcsname\relax\global\def\draftcontrol{0}
\fi

{\count255=\time\divide\count255 by 60
\xdef\hourmin{\number\count255}
\multiply\count255 by-60\advance\count255 by\time
\xdef\hourmin{\hourmin:\ifnum\count255<10 0\fi\the\count255}}
\def\draftdate{\number\month/\number\day/\number\year\ \ \ \hourmin }

\newcommand\makepapertitle{\par
  \begingroup
    \renewcommand\thefootnote{\@fnsymbol\c@footnote}%
    \def\@makefnmark{\rlap{\@textsuperscript{\normalfont\@thefnmark}}}%
    \long\def\@makefntext##1{\parindent 1em\noindent
            \hb@xt@1.8em{%
                \hss\@textsuperscript{\normalfont\@thefnmark}}##1}%
     \newpage
     \global\@topnum\z@   
     \@makepapertitle
     \thispagestyle{empty}\@thanks
  \endgroup
  \setcounter{footnote}{0}%
  \global\let\thanks\relax
  \global\let\makepapertitle\relax
  \global\let\@makepapertitle\relax
  \global\let\@thanks\@empty
  \global\let\@author\@empty
  \global\let\@date\@empty
  \global\let\@title\@empty
  \global\let\title\relax
  \global\let\author\relax
  \global\let\date\relax
  \global\let\and\relax
  \def\version{\let\version\@version\@gobble}
}
\def\@makepapertitle{%
  \newpage
   \ifnum\draftcontrol=1 {}
   \version\versionno
   \vskip 3em%
   \else
   \hfill\hbox to 3cm {\parbox{4cm}{\@pubnum}\hss}%
   \vskip 3em%
   \fi
   \begin{center}%
   \let \footnote \thanks
     {\LARGE {\@title}}%
     \vskip 1.5em%
     {\normalsize
       \lineskip .5em%
       \begin{tabular}[t]{c}%
         \@author
       \end{tabular}\par}%
     \vskip 1.5em%
     {\@bstract}%
     \end{center}%
     \vskip 1.5em
     \@date%
   \par
}

\gdef\@pubnum{}
\def\pubnum#1{%
  \gdef\@pubnum{#1}}

\gdef\@bstract{}
\def\Abstract#1{%
  \gdef\@bstract{%
   \parbox{\textwidth-0pc}{%
   \centerline{\bf Abstract}\penalty1000%
\kern.2cm%
\noindent
\renewcommand\baselinestretch{1.0}%
{#1}}}
}

\def\ps@paper{\let\@mkboth\@gobbletwo%
     \ifnum\draftcontrol=1
    \def\@oddfoot{\hbox to \textwidth{\tiny \versionno \hfil\tiny\draftdate}%
    \hskip -\textwidth \hbox to \textwidth{\hfil\rm\thepage\hfil}}%
     \else\def\@oddfoot{\hbox to \textwidth{\hfil\rm\thepage\hfil}}
     \fi
     \let\@evenfoot\@oddfoot
}

\def\body{\clearpage
          \pagestyle{paper}
    }

\def\@version#1{\ifnum\draftcontrol=1
\typeout{}\typeout{#1}\typeout{}
\vskip3mm\centerline{\hbox{\fbox{\normalsize{\tt DRAFT -- #1 -- }
                   {\draftdate}}}}\vskip3mm
\fi}
\let\version\@version
\long\def\eqlabel#1{\ifnum\draftcontrol=1
                    \tag@false  
                    \tag*{(\theequation) \hbox to -0.2cm{\hspace{0cm}\small{#1}\hss}}
                    \refstepcounter{equation}
                    \edef\@currentlabel{\theequation}
                    \ltx@label{#1}          
                    \else
                    \label{#1}
                    \fi
                    }
\let\st@bibitem\@bibitem
\let\st@lbibitem\@lbibitem
\ifnum\draftcontrol=1
  \def\@bibitem#1{%
    \st@bibitem{#1}\a@@label{#1}\ignorespaces}
  \def\@lbibitem[#1]#2{%
    \st@lbibitem[#1]{#2}\a@@label{#2}\ignorespaces}
  \def\a@@label#1{%
    \gdef\a@lab{\smash{\normalfont\small#1}}
    \ifvmode
      \if@inlabel
        \global\setbox\@labels\hbox{%
          \llap{\a@lab\let\a@lab\relax
                \kern\@totalleftmargin\kern\marginparsep}%
          \box\@labels}%
      \fi
    \fi}
\fi

\documentclass[12pt,letterpaper]{article}

\usepackage{amsmath,amssymb,array,calc,epsfig,rotating,bm}
\usepackage[sort]{cite}
\usepackage{graphicx}
\usepackage{psfrag,verbatim}

\ifnum\draftcontrol=1
\fi

\renewcommand\baselinestretch{1.25}
\renewcommand\section{\@startsection {section}{1}{\z@}%
                                   {-3.5ex \@plus -1ex \@minus -.2ex}%
                                   {2.3ex \@plus.2ex}%
                                   {\normalfont\large\bfseries}}
\renewcommand\subsection{\@startsection{subsection}{2}{\z@}%
                                   {-3.25ex\@plus -1ex \@minus -.2ex}%
                                   {1.5ex \@plus .2ex}%
                                   {\normalfont\normalsize\bfseries}}
\renewcommand\subsubsection{\@startsection{subsubsection}{3}{\z@}%
                                   {-3.25ex\@plus -1ex \@minus -.2ex}%
                                   {1.5ex \@plus .2ex}%
                                   {\normalfont\normalsize\it}}
\renewcommand\paragraph{\@startsection{paragraph}{4}{\z@}%
                                   {-3.25ex\@plus -1ex \@minus -.2ex}%
                                   {1.5ex \@plus .2ex}%
                                   {\normalfont\normalsize\bf}}

\numberwithin{equation}{section}

\def\revise#1       {\raisebox{-0em}{\rule{3pt}{1em}}%
                     \marginpar{\raisebox{.5em}{\vrule width3pt\
                     \vrule width0pt height 0pt depth0.5em
                     \hbox to 0cm{\hspace{0cm}{%
                     \parbox[t]{4em}{\raggedright\footnotesize{#1}}}\hss}}}}

\newcommand\nxt[1]  {\\\fnxt#1}
\newcommand{\ie}{{\it i.e.,}\ }

\def\cale         {{\cal E}}
\def\calf         {{\cal F}}

\def\calk         {{\cal K}}
\def\call         {{\cal L}}
\def\calm         {{\cal M}}
\def\caln         {{\cal N}}
\def\calo         {{\cal O}}
\def\calp         {{\cal P}}

\def\zet          {{\mathbb Z}}

\def\del          {\partial}

\def\tr           {\mathop{\rm Tr}}

\def\Im           {{\rm Im\hskip0.1em}}

\def\sqr#1#2{{\vcenter{\vbox{\hrule height.#2pt
 \hbox{\vrule width.#2pt height#1pt \kern#1pt
 \vrule width.#2pt}\hrule height.#2pt}}}}
\def\square{%
  \mathop{\mathchoice{\sqr{12}{15}}{\sqr{9}{12}}{\sqr{6.3}{9}}{\sqr{4.5}{9}}}}

\newcommand{\kk}{\mathfrak{q}}
\newcommand{\ww}{\mathfrak{w}}
\newcommand{\WW}{{\Omega}}

\def\a{\alpha}

\def\w{\omega}

\def\dd{\delta}
\def\e{\epsilon}
\def\c{\chi}

\def\g{\gamma}

\def\aa1{\phi}
\def\cc1{\psi}

\def\l{\lambda}

\def\Om{\Omega}
\def\om{\Omega}
\def\tf{\tilde{f}}
\def\th{\tilde{h}}
\def\tK{\tilde{K}}
\def\tk{\tilde{k}}
\def\tg{\tilde{g}}
\def\ra{\Longrightarrow}

\catcode`\@=12

\begin{document}


\title{\bf $\chi$SB in cascading gauge theory plasma}
\pubnum{UWO-TH-10/9}

\date{December 2010}

\author{
Alex Buchel\\[0.4cm]
\it Department of Applied Mathematics\\
\it University of Western Ontario\\
\it London, Ontario N6A 5B7, Canada\\[0.2cm]
\it Perimeter Institute for Theoretical Physics\\
\it Waterloo, Ontario N2J 2W9, Canada\\[0.2cm]
 }

\Abstract{$\caln=1$ supersymmetric $SU(K+P)\times SU(K)$ cascading gauge theory of 
Klebanov et.al \cite{kt,ks} undergoes a  first-order finite 
temperature confinement/deconfinement 
phase transition at $T_c=0.6141111(3) \Lambda$, where $\Lambda$ is the strong coupling 
scale of the theory.  The deconfined phase of the theory, with the unbroken chiral symmetry,
extends down to  $T_u=0.8749(0) T_c$, where it becomes perturbatively unstable due to the 
condensation of the hydrodynamic (sound) modes. We show that  at $T_{\c {\rm SB}} 
=0.882503(0) T_c\ >\  T_u$ the deconfined 
phase of the cascading plasma is perturbatively unstable towards 
development of the chiral symmetry breaking ($\c $SB) condensates. 
We present evidence that the ground state of the 
cascading plasma for $T<T_{\c{\rm SB}}$ can not be homogeneous and isotropic.
}

\makepapertitle

\body

\version\versionno
\tableofcontents

\section{Introduction and Summary}

Consider $\caln=1$ four-dimensional supersymmetric $SU(K+P)\times SU(K)$
gauge theory with two chiral superfields $A_1, A_2$ in the $(K+P,\overline{K})$
representation, and two fields $B_1, B_2$ in the $(\overline{K+P},K)$.
This gauge theory has two gauge couplings $g_1, g_2$ associated with 
two gauge group factors,  and a quartic 
superpotential
\begin{equation}
W\sim \tr \left(A_i B_j A_kB_\ell\right)\e^{ik}\e^{j\ell}\,.
\end{equation}
When $P=0$ above theory flows in the infrared to a 
superconformal fixed point, commonly referred to as Klebanov-Witten (KW) 
theory \cite{kw}. At the IR fixed point KW gauge theory is 
strongly coupled --- the superconformal symmetry together with 
$SU(2)\times SU(2)\times U(1)$ global symmetry of the theory implies 
that anomalous dimensions of chiral superfields $\gamma(A_i)=\gamma(B_i)=-\frac 14$, \ie non-perturbatively large.

When $P\ne 0$, conformal invariance of the above $SU(K+P)\times SU(K)$
gauge theory is broken. It is useful to consider an effective description 
of this theory at energy scale $\mu$ with perturbative couplings
$g_i(\mu)\ll 1$. It is straightforward to evaluate NSVZ beta-functions for 
the gauge couplings. One finds that while the sum of the gauge couplings 
does not run
\begin{equation}
\frac{d}{d\ln\mu}\left(\frac{\pi}{g_s}\equiv \frac{4\pi}{g_1^2(\mu)}+\frac{4\pi}{g_2^2(\mu)}\right)=0\,,
\eqlabel{sum}
\end{equation}
the difference between the two couplings is  
\begin{equation}
\frac{4\pi}{g_2^2(\mu)}-\frac{4\pi}{g_1^2(\mu)}\sim P \ \left[3+2(1-\g_{ij})\right]\ \ln\frac{\mu}{\Lambda}\,,
\eqlabel{diff}
\end{equation}
where $\Lambda$  is the strong coupling scale of the theory and $\g_{ij}$ is an anomalous dimension of operators $\tr A_i B_j$.
Given \eqref{diff} and \eqref{sum} it is clear that the effective weakly coupled description of $SU(K+P)\times SU(K)$ gauge theory 
can be valid only in a finite-width energy band centered about $\mu$ scale. Indeed, extending effective description both to the UV 
and to the  IR one necessarily encounters strong coupling in one or the other gauge group factor. As  explained 
in \cite{ks}, to extend the theory past the strongly coupled region(s) one must perform a Seiberg duality \cite{sd}. 
Turns out, in this gauge theory, a Seiberg duality transformation is a self-similarity transformation of the effective description 
so that $K\to K-P$ as one flows to the IR, or $K\to K+P$ as the energy increases. Thus, extension of the effective 
$SU(K+P)\times SU(K)$ description to all energy scales involves and infinite sequence - a {\it cascade } - of Seiberg dualities
where the rank of the gauge group is not constant along RG flow, but changes with energy according to \cite{b,k,aby2} 
\begin{equation}
K=K(\mu)\sim 2 P^2 \ln \frac \mu\Lambda\,, 
\eqlabel{effk}
\end{equation}
at least as $\mu\gg \Lambda$.
To see \eqref{effk}, note that the rank changes by $\Delta K\sim P$ as $P\Delta\left(\ln\frac\mu\Lambda\right)\sim 1$.
Although there are infinitely many duality cascade steps in the UV, there is only a finite number of duality transformations as one 
flows to the IR (from a given scale $\mu$). The space of vacua of a generic cascading gauge theory was studied in details in 
\cite{dks}. In the simplest case, when $K(\mu)$ is an integer multiple of $P$, the cascading gauge theory confines in the 
infrared with a spontaneous breaking of the chiral symmetry \cite{ks}. 

Effective description of the cascading gauge theory in the UV suggests
that it must be ultimately defined as a theory with an infinite number
of degrees of freedom. If so, an immediate concern is whether such a
theory is renormalizable as a four dimensional quantum field
theory, \ie whether a definite prescription can be made for the
computation of all gauge invariant correlation functions in the
theory.  As was pointed out in \cite{ks}, whenever $g_s K(\mu)\gg 1$,
the cascading gauge theory allows for a dual holographic
description \cite{juan,adscft} as type IIB supergravity on a warped
deformed conifold with fluxes. The duality is always valid in the UV
of the cascading gauge theory; if, in addition, $g_s P\gg 1$ the
holographic correspondence is valid in the IR as well. It was shown
in \cite{aby} that a cascading gauge theory {\it defined} by its
holographic dual as an RG flow of type IIB supergravity on a warped
deformed conifold with fluxes is holographically renormalizable as a
four dimensional quantum field theory.

In this paper we study the equilibrium properties of the cascading gauge theory at finite 
temperature\footnote{Hydrodynamics of the 
cascading gauge theory plasma was discussed in \cite{hyd1,hyd2,hyd3}.}.  At temperatures $T\gg \Lambda$ the cascading 
plasma is in the deconfined phase with an unbroken chiral symmetry \cite{b,kbh1,kbh2}. The temperature-dependent 
effective rank $K(T)$ of the cascading theory is large, compare to $P$ \cite{aby}:
\begin{equation}
\frac{K(T)}{P^2}=\frac 12 \ln \left(\frac{64\pi^4 }{81}\ \times\ \frac{s T }{ \Lambda^4}\right)\qquad \ra\qquad 
 \frac{K(T)}{P^2}\ \approx\ 2\ \ln \frac{T}{\Lambda}\,,\qquad T\gg \Lambda\,.
\eqlabel{kp2}
\end{equation}      
In \eqref{kp2} $s$ is the entropy density of the plasma at equilibrium.
To leading order at higher temperature\footnote{See \cite{hyd3} for the high-temperature expressions to order 
$\calo\left(\frac{P^8}{K(T)^4}\right)$.}, the pressure $\calp$ and the energy density $\cale$ are given by \cite{aby}
\begin{equation}
\begin{split}
\frac{\calp}{sT}=&\frac 14\left(1-\frac{P^2}{K(T)^2}+\calo\left(\frac{P^4}{K(T)^2}\right)\right)\,,\cr
\frac{\cale}{sT}=&\frac 34\left(1+\frac 13\ \frac{P^2}{K(T)^2}+\calo\left(\frac{P^4}{K(T)^2}\right)\right)\,.
\end{split}
\eqlabel{pe}
\end{equation}
In addition to stress-energy tensor, the equilibrium state of the cascading plasma is characterized 
by the expectation values of two dimension-4 operators: $\calo_4^{K_0}$ and $\calo_4^{p_0}$, 
a dimension-6 operator $\calo_6$, and a dimension-8 operator $\calo_8$ --- see \cite{aby} for 
details. As one reduces the temperature, the pressure of the cascading plasma decreases, ultimately turning 
negative below  $T_c=0.6141111(3) \Lambda$ \cite{kt3}. At this point, cascading plasma undergoes a first-order 
confinement/deconfinement first transition. As the transition occurs via nucleation of bubbles of the 
confined phase, it is non-perturbative. The deconfined phase of the cascading plasma remains as a metastable phase 
all the way down to   $T_u=0.8749(0) T_c$, at which point it joins a perturbatively unstable branch of the theory
with negative specific heat\footnote{Critical phenomena in the cascading plasma in the vicinity 
of $T_u$ was discussed in \cite{bp1}. Further analysis of the relevant critical universality class 
were performed in \cite{bp2}.}, see \cite{hyd3}.  

The deconfined phase of the cascading plasma extensively studied in \cite{aby,kt3,hyd3} does not spontaneously break chiral 
$U(1)$ symmetry. The latter is obvious by the absence of the expectation values for 
dimension-3 operators in the studied thermal states. On the other hand, the zero-temperature 
supersymmetric ground state of the theory spontaneously breaks chiral symmetry $U(1)\supset \zet_2$ \cite{ks}.  
The question we would like to address in this paper is  whether or not  spontaneous symmetry breaking occurs 
in the deconfined phase of the cascading plasma. We emphasize {\it spontaneous} symmetry breaking  as opposite 
to considering thermal states of the mass-deformed cascading gauge theory.
It is fairly straightforward to study 
mass-deformed cascading gauge theory. In the latter case, one introduces the mass terms 
\begin{equation}
\mu_i\equiv \frac{m_i}{\Lambda}\,,\qquad i=1,2\,,
\eqlabel{masses}
\end{equation}
for the gauginos
($\caln=1$ fermionic superpartners of $SU(K+P)\times SU(K)$ gauge bosons). These mass terms explicitly break both the supersymmetry 
and the chiral $U(1)$ symmetry. As we show  in section 5, it is straightforward to construct homogeneous and isotropic 
thermal states of the mass-deformed cascading plasma. Necessarily, these states have nonzero expectation value for 
dimension-3 operators 
\begin{equation}
\calo_3^{j}=\calo_3^{j}(\mu_i)\,,\qquad j=1,2\,,
\eqlabel{vevs3}
\end{equation}
 (gaugino bilinear condensates of the two gauge group factors). We show that in the chiral limit $\mu_i\to 0$,
the condensates vanish as well:
\begin{equation}
\lim_{\mu_i\to 0}\ \calo_3^{j}(\mu_i)=0\,.
\eqlabel{chirallimi}
\end{equation}  
Naively, the statement \eqref{chirallimi} would imply that the deconfined cascading plasma does not the break chiral
symmetry. We argue in section 3 that this is not the case. Specifically, we carefully study physical excitations in the 
cascading plasma, responsible for the development of the chiral condensates, and show that these fluctuations 
become tachyonic at temperatures $T<T_{\c\rm {SB}}=0.882503(0) T_c$. Thus, they must condense. The vanishing of the homogeneous condensates in
the chiral limit \eqref{chirallimi} strongly suggests that the 'chiral tachyons' we discover in section 3 
condense with a finite momentum --- the resulting ground state can not be homogeneous and isotropic.
We present a further (technical) evidence for the latter in section 4.

A more detailed outline of the rest of the paper follows. 
In section 2 we present consistent truncation of type IIB supergravity on warped deformed conifold with 
fluxes and $SU(2)\times SU(2)\times \zet_2$ global symmetry. The resulting five-dimensional effective gravitational 
action, which we  refer to as a 'KS effective action', is dual to a strongly coupled cascading gauge theory. 
In section 2.1 we discuss further truncation of the KS effective action to the Klebanov-Tseytlin (KT) effective action 
derived in \cite{aby}. In section 2.2 we derive equations of motion 
for the homogeneous and isotropic states of the cascading gauge theory (at zero or non-zero temperatures) and 
recover supersymmetric KT solution \cite{kt}, supersymmetric KS solution \cite{ks}, and the gravitational solutions 
describing the deconfined chirally symmetric phase of the cascading plasma \cite{aby,kt3}.  In section 2.3 
we derive the effective action for the linearized fluctuations dual to chiral condensates 
about chirally-symmetric states of the cascading theory. In 
section 3 we compute the spectrum of quasinormal modes of the chiral fluctuations described by the effective 
action of section 2.2 about the deconfined chirally-symmetric states of the cascading plasma. We show that 
for temperatures $T<T_{\c\rm {SB}}=0.882503(0) T_c$ these quasinormal modes realize Gregory-Laflamme instability of the 
translationary invariant Klebanov-Tseytlin horizons of \cite{kt3}. Since the deconfined cascading plasma is
thermodynamically stable down to $T_u$, and $T_{\c\rm {SB}}>T_u$, the gravitational dual to cascading 
gauge theory plasma presents an interesting string-theoretic example of violation of the correlated stability 
conjecture (CSC) \cite{gm1,gm2}\footnote{See \cite{bp3} for a recent discussion of CSC.}.  
In section 4 we discuss the gravitational solutions describing the homogeneous and isotropic states of the cascading plasma
with spontaneously broken chiral symmetry. We attempt (unsuccessfully) to construct these solutions by 
deforming chirally symmetric states of the cascading plasma for $T<T_{\c\rm{SB}}$ along the tachyonic directions 
discovered in section 3.  Finally, in section 5 we construct homogeneous and isotropic gravitational solutions 
dual to equilibrium states of mass-deformed cascading plasma for $T<T_{\c\rm{SB}}$. 
Constructed thermal states explicitly break chiral symmetry. We show that in the chiral limit these homogeneous and isotropic 
states do not break chiral symmetry spontaneously, see \eqref{chirallimi}.

\section{KS effective action}
We take a perspective of \cite{aby} where the cascading gauge theory at strong coupling is defined via its holographic 
dual, \ie by type IIB string theory on warped deformed conifold with fluxes and $SU(2)\times SU(2)\times \zet_2$
global symmetry. We begin with deriving an effective five-dimensional gravitational action representing the 
holographic dual of the cascading gauge theory.

We will work in the gravitational approximation to type IIB string theory,
using the type IIB supergravity action. This action takes the form (in 
the Einstein frame)
\begin{equation}
\begin{split}
S_{10}=\frac{1}{16\pi G_{10}}\int_{\calm_{10}}&\ \biggl(
R_{10}\wedge \star 1 -\frac 12 d\Phi\wedge \star d\Phi
-\frac 12 e^{-\Phi} H_3\wedge\star H_3  -\frac 12 
e^{\Phi} F_3\wedge\star F_3\\
& \qquad -\frac 14 F_5\wedge \star F_5-\frac 12 C_4\wedge H_3\wedge F_3\biggr),
\end{split}
\eqlabel{10action}
\end{equation} 
where $\calm_{10}$ is the ten dimensional bulk space-time, $G_{10}$ is
the ten dimensional gravitational constant, and we have consistently
set the axion $C_0$ to zero (it vanishes in all the solutions we
are interested in). In this action
\begin{equation}
F_3=dC_2\,,\qquad F_5=dC_4 -C_2\wedge H_3\,,
\eqlabel{fluxes}
\end{equation}
where $C_2$ and $C_4$ are the Ramond-Ramond (RR) potentials.
The equations of motion following from the 
action \eqref{10action} have to be supplemented by
the self-duality condition
\begin{equation}
\star F_5=F_5\,.
\eqlabel{5self}
\end{equation}
It is important to remember that the self-duality 
condition \eqref{5self} 
can not be imposed at the level of the action,
as this would lead to wrong equations of motion.

Introduce the following 1-forms on $T^{1,1}$ \cite{ks}:
\begin{equation}
\begin{split}
&g_1=\frac{\a^1-\a^3}{\sqrt 2}\,,\qquad g_2=\frac{\a^2-\a^4}{\sqrt 2}\,,\\
&g_3=\frac{\a^1+\a^3}{\sqrt 2}\,,\qquad g_4=\frac{\a^2+\a^4}{\sqrt 2}\,,\\
&g_5=\a^5\,,
\end{split}
\eqlabel{3form1}
\end{equation}
where 
\begin{equation}
\begin{split}
&\a^1=-\sin\theta_1 d\phi_1\,,\qquad \a^2=d\theta_1\,,\\
&\a^3=\cos\psi\sin\theta_2 d\phi_2-\sin\psi d\theta_2\,,\\
&\a^4=\sin\psi\sin\theta_2 d\phi_2+\cos\psi d\theta_2\,,\\
&\a^5=d\psi+\cos\theta_1 d\phi_1+\cos\theta_2 d\phi_2\,.
\end{split}
\eqlabel{3form2}
\end{equation}

The Einstein-frame metric ansatz is
\begin{equation}
ds_{10}^2 =g_{\mu\nu}(y) dy^{\mu}dy^{\nu}+\om_1^2(y) g_5^2
+\om_2^2(y) \left[g_3^2+g_4^2\right]+\om_3^2(y) \left[g_1^2+g_2^2\right],
\eqlabel{10met}
\end{equation}
where $y$ denotes the coordinates of $\calm_5$ (greek indices $\mu,\nu$
will run from $0$ to $4$).
Additionally, we assume the following ansatz for the 
fluxes $H_3\equiv d B_2$, $F_3$ and the 
dilaton $\Phi$ : 
\begin{equation}
\begin{split}
&B_2=h_1(y)\ g_1\wedge g_2+h_3(y)\ g_3\wedge g_4\,,\\
&F_3=\frac 19 P\ g_5\wedge g_3\wedge g_4+h_2(y)\ \left(g_1\wedge g_2-g_3\wedge g_4\right)\wedge g_5
\\
&\qquad +\left(g_1\wedge g_3+g_2\wedge g_4\right)\wedge d\left(h_2(y)\right)\,,\\
&\Phi= \Phi(y)\,,
\end{split}
\eqlabel{fdil}
\end{equation}
where 
$P$ is an integer corresponding to the RR 3-form flux on the compact 3-cycle
(and to the number of fractional branes on the conifold).
Special care should be taken with the RR 5-form. 
From \eqref{fluxes} we get the Bianchi identity
\begin{equation}
dF_5=-F_3\wedge H_3\,,
\eqlabel{5form}
\end{equation}
which for the background fluxes \eqref{fdil}
is solved by
\begin{equation}
F_5=dC_4+\bigg(4{\om}_0+ h_2(y)\left(h_3(y)-h_1(y)\right)+\frac 19 P h_1(y)\bigg)\ 
g_5 \wedge
g_3\wedge g_4\wedge 
g_1\wedge g_2\,,
\eqlabel{f5sol}
\end{equation}
with some constant ${\om}_0$.
In our ansatz the RR four-form does not 
depend on the compact coordinates, that is $C_4\equiv C_4(y)$ (note that
$C_4\wedge F_3\wedge H_3\ne 0$), and the RR five-form is proportional to
the volume form of $\calm_5$ (plus its dual). We define $F(y)$ by
\begin{equation}
dC_4=\frac{F(y)}{\Omega_1 \Omega_2^2\om_3^2}
\ {\rm vol}_{\calm_5}\equiv \frac{F(y)}{\Omega_1 \Omega_2^2\om_3^2}
\sqrt{-\det(g_{\mu\nu})}\ 
dy^{1}\wedge\cdots
\wedge dy^5\,,
\eqlabel{dc4}
\end{equation} 
and then the self-duality condition \eqref{5self}  implies 
\begin{equation}
F(y)=4{\om}_0+ h_2(y)\left(h_3(y)-h_1(y)\right)+\frac 19 P h_1(y)\,,
\eqlabel{kdef}
\end{equation}
(again, in deriving the effective action we should keep $C_4$  
unconstrained and impose this equation later). 
Altogether, from the five-dimensional perspective we allow fluctuations 
in the metric $g_{\mu\nu}(y)$, in the scalar fields
$\Omega_1(y)\,, \Omega_2(y)\,, \om_3(y)\,, h_1(y)\,, h_2(y)\,, h_3(y)\,,  \Phi(y)$
and in the four-form $C_4(y)$ (which is determined in terms of the others
by the self-duality condition). We have set to zero various fluctuations
of the form fields which are $p$-forms on $\calm_5$, and also fluctuations
of $C_2$ of the same form as the fluctuation of $B_2$ in \eqref{fdil},
even though they are
allowed by the symmetries. This is a consistent truncation of the
full ten dimensional supergravity action.

We now perform the KK reduction of \eqref{10action} by plugging into
it the ansatz described above. Recall that 
\begin{equation}
{\rm vol}_{T^{1,1}}\equiv \frac{1}{108} \int g_5 \wedge
g_3\wedge g_4\wedge 
g_1\wedge g_2=\frac{1}{108}\times \frac {16 \pi^3}{27}\,.
\end{equation}
First, we have
\begin{equation}
\int_{\calm_{10}} 1\wedge \star 1= 108\
{\rm vol}_{T^{1,1}}\int_{\calm_5} \Omega_2\Omega_2^2\om_3^2\ {\rm vol}_{\calm_5}\,.
\eqlabel{int5}
\end{equation}
With a straightforward but somewhat tedious computation 
we find that in the background \eqref{10met}
\begin{equation}
\begin{split}
R_{10}=R_5&+\left(\frac{1}{2\om_1^2}+\frac{2}{\om_2^2}+\frac{2}{\om_3^2}-\frac{\om_2^2}{4\om_1^2\om_3^2}
-\frac{\om_3^2}{4\om_1^2\om_2^2}-\frac{\om_1^2}{\om_2^2\om_3^2}\right)-2\Box \ln\left(\om_1\om_2^2\om_3^2\right)\\
&-\biggl\{\left(\nabla\ln\om_1\right)^2+2\left(\nabla\ln\om_2\right)^2
+2\left(\nabla\ln\om_3\right)^2+\left(\nabla\ln\left(\om_1\om_2^2\om_3^2\right)\right)^2\biggr\}\,,
\end{split}
\eqlabel{ric5}
\end{equation}
where $R_5$ is the five dimensional Ricci scalar of the metric 
\begin{equation}
ds_{5}^2 =g_{\mu\nu}(y) dy^{\mu}dy^{\nu}\,.
\eqlabel{5met}
\end{equation}
In \eqref{ric5}, $\nabla_\lambda$ denotes the covariant derivative 
with respect to the metric \eqref{5met}, explicitly given by
\begin{equation}
\begin{split}
\nabla_\lambda \Omega_i &= \del_\lambda \Omega_i\,,\\
\nabla_\lambda\nabla_\nu \Omega_i &= \del_\lambda\del_\nu \Omega_i
-\Gamma^{\rho} _{\lambda\nu}\ \del_\rho \Omega_i\,.
\end{split}
\eqlabel{remcov}
\end{equation}
Now, by plugging our ansatz into \eqref{10action} we find that it
reduces to the following 
effective action : 
\begin{equation}
\begin{split}
S_5=& \frac{108}{16\pi G_5} \int_{\calm_5} {\rm vol}_{\calm_5}\ \Omega_1 \Omega_2^2\om_3^2\ 
\biggl\lbrace 
 R_{10}-\frac 12 \left(\nabla \Phi\right)^2\\
&-\frac 12 e^{-\Phi}\left(\frac{(h_1-h_3)^2}{2\om_1^2\om_2^2\om_3^2}+\frac{1}{\om_3^4}\left(\nabla h_1\right)^2
+\frac{1}{\om_2^4}\left(\nabla h_3\right)^2\right)
\\
&-\frac 12 e^{\Phi}\left(\frac{2}{\om_2^2\om_3^2}\left(\nabla h_2\right)^2
+\frac{1}{\om_1^2\om_2^4}\left(h_2-\frac P9\right)^2
+\frac{1}{\om_1^2\om_3^4} h_2^2\right)
\\
&-\frac 14\biggl(\frac{F^2}{\Omega_1^2\Omega_2^4\om_3^4}
+\frac{5}{24}\  \calf_{\mu_1\cdots\mu_5}
\calf^{\mu_1\cdots\mu_5}\biggr)
\biggr\rbrace\\
&+\frac{108}{16\pi G_5}\ \frac 12 \int_{\calm_5}\  dF\wedge C_4\,,
\end{split}
\eqlabel{5actionzero}
\end{equation}
where 
\begin{equation}
\calf_{\mu_1\cdots\mu_5}\equiv \del\ _{[\mu_1} C_4\ _{\mu_2\cdots \mu_5]}
=\frac 15 \frac{F}{\Omega_1 \Omega_2^2\om_3^2} 
\sqrt{-\det(g_{\mu\nu})}\ \epsilon_{\mu_1\cdots\mu_5}\,,
\eqlabel{defff}
\end{equation}
($[\cdots]$ denotes anti-symmetrization with 
weight one) 
and $G_5$ is the five dimensional effective gravitational constant  
\begin{equation}
G_5\equiv \frac{G_{10}}{{\rm vol}_{T^{1,1}}}\,.
\eqlabel{g5deff}
\end{equation}
Note that our gravitational action is not the standard five dimensional
action because of the factor of $\Omega_1 \Omega_2^2\om_3^2$ in front of
the five dimensional Einstein-Hilbert term.

In the five dimensional action it turns out to be possible to ``integrate
out'' the field $C_4$ using the self-duality equation \eqref{kdef} and
to obtain an action involving only the other fields. This leads to the
action we will be using in this paper
\begin{equation}
\begin{split}
S_5=& \frac{108}{16\pi G_5} \int_{\calm_5} {\rm vol}_{\calm_5}\ \Omega_1 \Omega_2^2\om_3^2\ 
\biggl\lbrace 
 R_{10}-\frac 12 \left(\nabla \Phi\right)^2\\
&-\frac 12 e^{-\Phi}\left(\frac{(h_1-h_3)^2}{2\om_1^2\om_2^2\om_3^2}+\frac{1}{\om_3^4}\left(\nabla h_1\right)^2
+\frac{1}{\om_2^4}\left(\nabla h_3\right)^2\right)
\\
&-\frac 12 e^{\Phi}\left(\frac{2}{\om_2^2\om_3^2}\left(\nabla h_2\right)^2
+\frac{1}{\om_1^2\om_2^4}\left(h_2-\frac P9\right)^2
+\frac{1}{\om_1^2\om_3^4} h_2^2\right)
\\
&-\frac {1}{2\Omega_1^2\Omega_2^4\om_3^4}\left(4{\om}_0+ h_2\left(h_3-h_1\right)+\frac 19 P h_1\right)^2
\biggr\rbrace\,,\\
\end{split}
\eqlabel{5action}
\end{equation}
where $R_{10}$ is given by \eqref{ric5}.

\subsection{Reduction of KS effective action to KT effective action}
Effective action \eqref{5action} allows for further consistent truncation. Indeed, 
setting 
\begin{equation}
\begin{split}
&ds_{5}^2=\left(ds_{5}^2\right)^{KT}\,,\qquad \om_1=\frac{1}{3}{{\om}_1^{KT}}\,,\qquad \om_2=\om_3=\frac{1}{\sqrt{6}}{\om}^{KT}\,,\qquad  
F=\frac{K^{KT}}{108}\,,\\
&h_1=h_3=\frac 16\ \tilde{k}^{KT}\,,\qquad h_2=\frac{P}{18}\,,\qquad \Phi=\Phi^{KT}\,,\qquad \Omega_0=\frac{\tilde{K}_0^{KT}}{432}\,,
\end{split}
\eqlabel{ksktred}
\end{equation}
we obtain effective action of \cite{aby} describing $SU(2)\times SU(2)\times U(1)$ symmetric states of the 
cascading gauge theory.
In \eqref{ksktred} we used superscript 'KT' to relate to fields of the KT effective action in \cite{aby}.

\subsection{Homogeneous and isotropic $SU(2)\times SU(2)\times \zet_2$ states of the cascading gauge theory }
In this section we derive gravitational equations of motion from the effective 
action \eqref{5action} describing homogeneous and isotropic states of the cascading gauge theory
at zero and nonzero temperature. In the latter case the background geometry has a regular 
(homogeneous and isotropic) Schwartzchild horizon. We recover from the obtained equations 
of motion supersymmetric Klebanov-Tseytlin \cite{ks} and Klebanov-Strassler \cite{ks}
solutions, as well as KT BH solution of \cite{kt3}.

The general five-dimensional background geometry with homogeneous and isotropic (but not necessary
Lorentz-invariant) asymptotic boundary takes form
\begin{equation}
\begin{split}
ds_{5}^2=&H^{-1/2}\left(-f_1^2 dt^2 +dx_1^2+dx_2^2+dx_3^2\right)+H^{1/2}\w_1^2\ \frac{dr^2}{\tf_2^2}\,,\\
\Omega_i=&\w_i H^{1/4}\,,\qquad  g_s= e^\Phi\,,
\end{split}
\eqlabel{metricks}
\end{equation}
where $H=H(r)$, $f_1=f_1(r)$, $\tf_2=\tf_2(r)$ and $\w_i=\w_i(r)$. Additionally, we set $h_i=h_i(r)$ 
and $g_s=g_s(r)$. 

From \eqref{5action} we find the following equations of motion\footnote{We verified that exactly the same equations of 
motion arise directly from type IIB supergravity in ten dimensions.}
\begin{equation}
\begin{split}
0=&h_1''+h_1' \left[\ln\frac{f_1 \tf_2 \omega_2^2}{\omega_3^2 H g_s}\right]'-h_1 \biggl( \frac{g_s (P-9 h_2)^2}
{81H \tf_2^2 \omega_2^4}+\frac{\omega_3^2}{2\tf_2^2 \omega_2^2}\biggr)+\frac{(h_2 h_3+4 \Omega_0) (9h_2-P)g_s}{9H \tf_2^2 \omega_2^4}\\
&+\frac{\omega_3^2 h_3}{2\omega_2^2 \tf_2^2}\,,
\end{split}
\eqlabel{ks1}
\end{equation}
\begin{equation}
\begin{split}
0=&h_2''+h_2' \left[\ln\frac{f_1 \tf_2 g_s}{H}\right]'-h_2 \biggl( \frac{\omega_2^4+\omega_3^4}{2\tf_2^2 \omega_3^2 \omega_2^2}
+\frac{(h_1-h_3)^2}{2H \tf_2^2 \omega_2^2 \omega_3^2 g_s}\biggr)+\frac{(h_1-h_3)(P h_1+36 \Omega_0)}{18H \tf_2^2 \omega_2^2 \omega_3^2 g_s}
\\
&+\frac{\omega_3^2 P}{18\tf_2^2 \omega_2^2}\,,
\end{split}
\eqlabel{ks2}
\end{equation}
\begin{equation}
\begin{split}
0=&h_3''+h_3' \left[\ln\frac{f_1 \tf_2 \omega_3^2}{\omega_2^2 H g_s}\right]' 
-\frac{h_3(2 h_2^2 g_s+\omega_2^2 H \omega_3^2)}{2H \tf_2^2 \omega_3^4}+\frac{g_s (h_1 h_2-4 \Omega_0) (9h_2-P)}
{9H \tf_2^2 \omega_3^4}+\frac{\omega_2^2 h_1}{2\tf_2^2 \omega_3^2}\\
&-\frac{4g_s \Omega_0 P}{9H \tf_2^2 \omega_3^4}\,,
\end{split}
\eqlabel{ks3}
\end{equation}
\begin{equation}
\begin{split}
0=&g_s''-\frac{(g_s')^2}{g_s}+g_s' \left[\ln{f_1 \tf_2} {\omega_2^2 \omega_3^2}\right]'-\frac{g_s^2}{H} \biggl(
\frac{(h_2')^2}{\omega_2^2  \omega_3^2} +\frac{h_2^2}{2\tf_2^2  \omega_3^4}+\frac{(9 h_2-P)^2}{162 \tf_2^2 \omega_2^4 }\biggr)
\\
&+\frac{1}{2H} \biggl(\frac{(h_1')^2}{\omega_3^4}+\frac{(h_3')^2}{\omega_2^4}\biggr)
+  \frac{(h_3-h_1)^2}{4\omega_2^2 \omega_3^2 \tf_2^2 H}\,,
\end{split}
\eqlabel{ks4}
\end{equation}
\begin{equation}
\begin{split}
0=&f_1''+f_1' \left[\ln \tf_2 \omega_2^2 \omega_3^2\right]'\,,
\end{split}
\eqlabel{ks5}
\end{equation}
\begin{equation}
\begin{split}
0=&H''-\frac{(H')^2}{H}+H' \left[\ln f_1 \tf_2 \omega_2^2 \omega_3^2\right]'
+\frac{(9 h_2 (h_3- h_1)+P h_1+36 \Omega_0)^2}{81\tf_2^2 \omega_2^4 \omega_3^4 H}\\
&+\frac{1}{2g_s} \left(\frac{(h_1')^2}{\omega_3^4}+\frac{(h_3')^2}{\omega_2^4}\right)+ \frac{g_s(h_2')^2}{\omega_3^2 \omega_2^2} 
+ \frac{(h_3-h_1)^2}{4\tf_2^2 \omega_2^2 \omega_3^2 g_s}+\frac{g_s h_2^2}{2\tf_2^2 \omega_3^4}+
\frac{g_s (9 h_2-P)^2}{162\tf_2^2 \omega_2^4}\,,
\end{split}
\eqlabel{ks6}
\end{equation}
\begin{equation}
\begin{split}
0=&\omega_1''-\frac{(\omega_1')^2}{\omega_1}
+\omega_1' \left[\ln f_1 \tf_2 \omega_2^2 \omega_3^2\right]'-\frac{\omega_1}{4 H g_s} 
\left(\frac{(h_1')^2}{\omega_3^4}+\frac{(h_3')^2}{\omega_2^4}\right)- \frac{\omega_1 g_s(h_2')^2}{2\omega_3^2 \omega_2^2 H} 
\\
&+  \frac{\omega_1 ((\omega_2^2-\omega_3^2)^2-4 \omega_1^4)}{4\omega_2^2 \omega_3^2 \tf_2^2}
+\frac{\omega_1}{H} \biggl( \frac{(h_3-h_1)^2}{8\tf_2^2 \omega_2^2 \omega_3^2 g_s}+\frac{g_s h_2^2}{4\omega_3^4 \tf_2^2}
+\frac{g_s (9 h_2-P)^2}{324\tf_2^2 \omega_2^4}\biggr)\,,
\end{split}
\eqlabel{ks7}
\end{equation}
\begin{equation}
\begin{split}
0=&\omega_2''+\frac{(\omega_2')^2}{\omega_2}+\omega_2' \left[\ln f_1 \tf_2 \omega_3^2\right]'
-\frac{\omega_2}{4 H g_s} \left(\frac{(h_1')^2}{\omega_3^4}-\frac{(h_3')^2}{\omega_2^4}\right) 
\\
&-\frac{g_s (81 h_2^2 \omega_2^4 - \omega_3^4(9 h_2-P)^2)}{324
\omega_2^3 \omega_3^4 \tf_2^2 H} 
-\frac{(\omega_2^4-\omega_3^4+8 \omega_3^2 \omega_1^2-4 \omega_1^4)}{8\omega_3^2 \omega_2 \tf_2^2}\,,
\end{split}
\eqlabel{ks8}
\end{equation}
\begin{equation}
\begin{split}
0=&\omega_3''+\frac{(\omega_3')^2}{\omega_3}+\omega_3' \left[\ln f_1 \tf_2 \omega_2^2\right]'
+\frac{\omega_3}{4 H g_s} \left(\frac{(h_1')^2}{\omega_3^4}-\frac{(h_3')^2}{\omega_2^4}\right) \\
&+\frac{g_s (81 h_2^2\omega_2^4 -\omega_3^4 (9h_2-P)^2 )}
{324\omega_3^3 \omega_2^4 \tf_2^2 H}-\frac{\omega_3^4-\omega_2^4+8 \omega_2^2 \omega_1^2-4 \omega_1^4}{
8\omega_3 \omega_2^2 \tf_2^2}\,,
\end{split}
\eqlabel{ks9}
\end{equation}
\begin{equation}
\begin{split}
0=&\left(\left[\ln g_s\right]'\right)^2+\left(\left[\ln H\right]'\right)^2+\frac{1}{H g_s} 
\left(\frac{(h_1')^2}{\omega_3^4}+\frac{(h_3')^2}{\omega_2^4}\right)+2 \frac{g_s(h_2')^2}{H \omega_3^2 \omega_2^2}
-4 \left(\left[\ln \omega_2 \omega_3\right]'\right)^2\\
&-8 \left[\ln \omega_3\right]' \left[\ln \omega_1 \omega_2 f_1\right]'
-\left[\ln f_1\right]' \left[\ln H^2 \omega_1^4 \omega_2^8\right]'-8 \left[\ln\omega_1\right]' \left[\ln\omega_2\right]'
\\
&-\frac{(9 h_2 (h_3-h_1)+P h_1+36 \Omega_0)^2}{81\omega_2^4 \tf_2^2 \omega_3^4 H^2} 
-\frac{(\omega_2^2-\omega_3^2)^2-4\omega_1^2(2\omega_2^2-\omega_1^2+2\omega_3^2)}{2\omega_3^2 \omega_2^2 \tf_2^2} \\
&-\frac{1}{H \tf_2^2} \left(
\frac{(h_1-h_3)^2}{2\omega_2^2 g_s \omega_3^2}+\frac{g_s (9 h_2-P)^2}{81\omega_2^4}+\frac{g_s h_2^2}{\omega_3^4}\right)\,.
\end{split}
\eqlabel{ksc}
\end{equation}

We explicitly verified that the constraint \eqref{ksc} associated with the reparametrization of the radial coordinate $r$
is consistent with the second order equations of motion \eqref{ks1}-\eqref{ks9}. 

\subsubsection{Supersymmetric KT solution}
The singular KT solution \cite{kt} to \eqref{ks1}-\eqref{ksc} is given by 
\begin{equation}
\begin{split}
&h_1=h_3=-\frac 16 P g_0\ \ln r\,,\qquad h_2=\frac{P}{18}\,,\qquad 
f_1=1\,, \qquad \tf_2=\frac r3\,,\qquad g_s=g_0\\
&\w_1=\frac{1}{3r}\,,\qquad \w_2=\w_3=\frac{1}{\sqrt{6}r}\,,\qquad
H=r^4\left(108\Om_0+\frac 18P^2 g_0-\frac 12 P^2g_0 \ln r\right)
\end{split}
\eqlabel{kskt}
\end{equation}
where $r\to 0$ is the boundary.

\subsubsection{Supersymmetric KS solution}
The supersymmetric KS solution \cite{ks} to \eqref{ks1}-\eqref{ksc} is given by 
\begin{equation}
\begin{split}
&h_1=\frac{Pg_0(\cosh r-1)}{18\sinh r}
\left(\frac{r\cosh r}{\sinh r}-1\right)\,,\qquad 
h_2=\frac{P}{18}\left(1-\frac {r}{\sinh r}\right)\,,\\
&h_3=\frac{Pg_0(\cosh r+1)}{18\sinh r}
\left(\frac{r\cosh r}{\sinh r}-1\right)\,,\qquad f_1=\tf_2=1\,,
\qquad g_s=g_0\,,\\
&\w_1=\frac{\epsilon^{2/3}}{\sqrt{6}{\hat K}}\,,\qquad 
\w_2=\frac{\epsilon^{2/3}{\hat K}^{1/2}}{\sqrt{2}}\cosh\frac r2\,,\qquad  \w_3=\frac{\epsilon^{2/3}{\hat K}^{1/2}}{\sqrt{2}}
\sinh\frac r2\,,
\end{split}
\eqlabel{ksks}
\end{equation}
with 
\begin{equation}
{\hat K}=\frac{(\sinh (2r)-2r)^{1/3} }{2^{1/3}\sinh r}\,,\qquad H'=\frac{16((9 h_2-P)h_1-9 h_3 h_2)}{9\epsilon^{8/3}{\hat K}^2\sinh^2 r }\,,
\qquad \Om_0=0\,,
\eqlabel{kk}
\end{equation}
where now $r\to \infty$ is the boundary.

\subsubsection{KT BH solution}
The  KT BH equations of motion in the parametrization of \cite{kt3} are  obtained
from \eqref{ks1}-\eqref{ksc} introducing a radial coordinate
\begin{equation}
x\equiv 1-f_1(r)\,,
\eqlabel{xks}
\end{equation}
and setting 
\begin{equation}
\begin{split}
&h_1=h_3=\frac 1P\left(\frac {K}{12}-36\Om_0\right)\,,\qquad h_2(x)=\frac{P}{18}\,,  \qquad g_s(x)=g\,,\\
&\w_1(x)=\frac{f_2^{1/2}}{3(2x-x^2)^{1/4}}\,, \qquad \w_2=\w_3= \frac{f_3^{1/2}}{\sqrt{6}(2x-x^2)^{1/4}}\,,\\
&H(x)= (2x-x^2) h\,.
\end{split}
\eqlabel{ksktbh}
\end{equation}

\subsection{Chiral fluctuations in cascading plasma}
Recall that unlike \eqref{5action}, the effective action obtained with further 
consistent truncation 
\begin{equation}
h_1=h_3\,,\qquad h_2=\frac{P}{18}\,,\qquad \Om_2=\Om_3\,,
\eqlabel{further}
\end{equation}
has an enlarged global symmetry, \ie $\zet_2$ get enhanced to $U(1)$.
On the dual gauge theory side such enhancement corresponds to restoration 
of the chiral symmetry. As familiar from \cite{ks}, the chiral symmetry of the 
cascading theory is spontaneously broken at a supersymmetric ground state. 
In this  section we compute effective gravitational action of the 
linearized fluctuations about chirally symmetric states of the 
cascading gauge theory.
 
We introduce 
\begin{equation}
\begin{split}
h_1=&\frac 1P\left(\frac{K_1}{12}-36\Om_0\right)\,,\qquad h_2=\frac{P}{18}\ K_2\,,\qquad 
h_3=\frac 1P\left(\frac{K_3}{12}-36\Om_0\right)\,,\\
\Om_1=&\frac 13 f_c^{1/2} h^{1/4}\,,\qquad \Om_2=\frac {1}{\sqrt{6}} f_a^{1/2} h^{1/4}\,,\qquad 
\Om_3=\frac {1}{\sqrt{6}} f_b^{1/2} h^{1/4}\,.
\end{split}
\eqlabel{redef}
\end{equation}
In it straightforward to verify that linearized fluctuations $\{\dd f,\dd k_1,\dd k_2\}$ in
\begin{equation}
\begin{split}
K_1=&K+\dd k_1\,,\qquad K_2=1+\dd k_2\,,\qquad K_3=K-\dd k_1\,,
\\
f_c=&f_2\,,\qquad f_a=f_3+\dd f\,,\qquad f_b=f_3-\dd f\,,
\end{split}
\eqlabel{deffl}
\end{equation}
decouple from all the other fluctuations, provided the gravitational fields 
\begin{equation}
\bigg\{ds_5^2\,, K\,, h\,, 
f_2\,, f_3\,, g_s\bigg\} 
\eqlabel{symmetric}
\end{equation}
are on-shell, \ie describe  a chirally symmetric state of the
cascading plasma. The effective action for the $\c$SB fluctuations can be derived from \eqref{5action}:
\begin{equation}
S_{\c\rm{SB}}\bigg[\dd f,\dd k_1,\dd k_2\bigg]
=\frac{1}{16\pi G_5}\int_{\calm_5}\ {\rm vol}_{\calm_5}\ h^{5/4}f_2^{1/2}f_3^2
\biggl\{\call_1+\call_2+\call_3+\call_4+\call_5\biggr\}\,,
\eqlabel{flaction}
\end{equation}
\begin{equation}
\begin{split}
\call_1=&-\frac{(\dd f)^2}{f_3^2}\left(
-\frac{P^2 e^\Phi}{2 f_2 h^{3/2} f_3^2}-\frac{(\nabla K)^2}{8 f_3^2 h P^2 e^\Phi}- \frac{K^2}{2f_2 h^{5/2} f_3^4}
\right)\,,
\end{split}
\eqlabel{call1}
\end{equation}
\begin{equation}
\begin{split}
\call_2=&-\frac{9f_3^2-24 f_2 f_3+4f_2^2}{f_2h^{1/2}f_3^4}\ (\dd f)^2+2\square\frac{(\dd f)^2}{f_3^2}
-\left(\nabla \frac{(\dd f)^2}{f_3^2}\right)^2\\
&-2\nabla\left(\ln h^{1/4}f_3^{1/2}\right)\nabla 
\left(\frac{(\dd f)^2}{f_3^2}\right)+2\nabla\left(\ln f_2^{1/2} h^{5/4}f_3^2\right)\nabla\left(\frac{(\dd f)^2}
{f_3^2}\right)\,,
\end{split}
\eqlabel{call2}
\end{equation}
\begin{equation}
\begin{split}
\call_3=&-\frac {1}{2P^2 e^\Phi}\biggl(\frac {9}{2 f_2 h^{3/2}f_3^2}\ (\dd k_1)^2
+\frac {1}{2h f_3^4} \biggl(2(\nabla K)^2\ (\dd f)^2+f_3^2\ (\nabla\dd k_1)^2
\\
&+4f_3 \dd f\ \nabla K\nabla \dd k_1
\biggr)
\biggr)\,,
\end{split}
\eqlabel{call3}
\end{equation}
\begin{equation}
\begin{split}
\call_4=&\frac{P^2 e^\Phi}{2}\biggl(\frac {2}{9hf_3^2}\ (\nabla \dd k_2)^2
+\frac{2}{f_2h^{3/2}f_3^4}\left(3\ (\dd f)^2+4f_3\ \dd f\dd k_2+f_3^3\ (\dd k_2)^2 \right)
\biggr)\,,
\end{split}
\eqlabel{call4}
\end{equation}
\begin{equation}
\begin{split}
\call_5=&\frac{K}{f_2 h^{5/2} f_3^6}\ \left(f_3^2\ \dd k_1\dd k_2-K\ (\dd f)^2\right)\,.
\end{split}
\eqlabel{call5}
\end{equation}

\section{$\c$SB quasinormal modes of the KT BH}
In this section we study the spectrum of the $\c$SB  quasinormal modes of Klebanov-Tseytlin
black hole \cite{kt3,hyd3} and show that these modes are unstable (tachyonic) once 
$T<T_{\c\rm{SB}}$, with
\begin{equation}
T_{\c\rm{SB}}=0.882503(0)\ T_c\,,
\eqlabel{tscb}
\end{equation}
where $T_c$ is the critical temperature of the first-order confinement deconfinement 
phase transition. Although the spontaneous breaking of the chiral symmetry 
occurs in the meta-stable deconfined phase of the cascading plasma, this perturbative instability 
precedes the hydrodynamic instability in of the deconfined phase discovered in \cite{hyd3}
since $T_{\c\rm{SB}}>T_{u}=0.8749(0) T_c$. 

Effective action describing the $\c$SB fluctuations in cascading plasma is given by 
\eqref{flaction}-\eqref{call5}. The background geometry  dual to the deconfined 
homogeneous and isotropic phase of the cascading plasma is given by (see \eqref{metricks}
and \eqref{redef})
\begin{equation}
ds_5^2=h^{-1/2}(1-f_1^2)^{-1/2}\left(-f_1^2\ dt^2+dx_1^2+dx_2^2+dx_3^2\right)+\frac 13 h^{1/2}f_2\
\frac{dr^2}{\tf_2^2} \,,
\eqlabel{background}
\end{equation} 
with $\{f_1,\tf_2,K,h,f_2,f_3,g_s\}$ being functions of $r$ only.
Without loss of generality we assume 
\begin{equation}
\dd f=e^{-i\w t+i k x_3} F\,,\qquad \dd k_1=e^{-i\w t+i k x_3} \calk_1\,,\qquad \dd k_2=e^{-i\w t+i k x_3} \calk_2\,,
\eqlabel{flucw}
\end{equation}
where  $\{F,\calk_1,\calk_2\}$ are functions of the radial coordinate only, satisfying the following 
equations of motion
\begin{equation}
\begin{split}
0=&-\w^2\ F+f_1^2 k^2\ F-\frac{9 \tf_2^2 f_1^2 }{h (1-f_1^2)^{1/2} f_2}\ F''\\
&+ \frac{ 9\tf_2 f_1}{h (1-f_1^2)^{3/2} f_2}(f_1^3 \tf_2'-\tf_2 f_1' f_1^2-f_1 \tf_2'-\tf_2 f_1')\ F'
\\
&+\frac{9f_1^2 K' \tf_2^2 }{2(1-f_1^2)^{1/2} h^2 g_s f_2 f_3}\ \calk_1'
-\frac{f_1^2} {2(1-f_1^2)^{5/2} f_2 g_s h^2 f_3^2}\biggl(-18 f_3^2 h g_s f_1^4-4 g_s^2 f_1^4\\
&+18 h g_s \tf_2^2 f_1^4 (f_3')^2-9 \tf_2^2 f_1^4 (K')^2+8 f_1^4 h g_s f_2^2+36 f_3^2 h g_s f_1^2+8 g_s^2 f_1^2
\\
&-36 h g_s \tf_2^2 f_1^2 (f_3')^2+18 \tf_2^2 f_1^2 (K')^2-16 f_1^2 h g_s f_2^2+8 h g_s f_2^2-4 g_s^2
+36 f_3^2 h g_s \tf_2^2 (f_1')^2\\
&-9 \tf_2^2 (K')^2+18 h g_s \tf_2^2 (f_3')^2-18 f_3^2 h g_s\biggr)\ F
+\frac{2  f_1^2 g_s}{(1-f_1^2)^{1/2} h^2 f_2 f_3}\ \calk_2\,,
\end{split}
\eqlabel{fequa}
\end{equation}
\begin{equation}
\begin{split}
0=&-\w^2\ \calk_1+f_1^2 k^2\ \calk_1-\frac{9 \tf_2^2 f_1^2} {h (1-f_1^2)^{1/2} f_2}\ \calk_1''\\
&-\frac{9 f_1 \tf_2}{g_s (1-f_1^2)^{3/2} f_2 h^2}
\biggl(h \tf_2 f_1^3 g_s'+\tf_2 f_1^3 g_s h'-h \tf_2 f_1 g_s'-\tf_2 f_1 g_s h'-f_1^3 g_s h \tf_2'
\\
&+\tf_2 f_1^2 g_s h f_1'+f_1 g_s h \tf_2'+\tf_2 g_s h f_1'\biggr)\  \calk_1'
-\frac{18 K' \tf_2^2 f_1^2 }{f_2 (1-f_1^2)^{1/2} h f_3}\ F'
\\
&+\frac{2 f_1^2}{h^2 f_2 f_3^3 (1-f_1^2)^{1/2}} \biggl(9 \tf_2^2 f_3' h K' f_3-2 g_s K\biggr)\ F
+\frac{(9 f_3^2 h \calk_1-2 g_s \calk_2 K) f_1^2}{h^2 f_2 (1-f_1^2)^{1/2} f_3^2}\,,
\end{split}
\eqlabel{k1equa}
\end{equation}
\begin{equation}
\begin{split}
0=&-\w^2\ \calk_2+f_1^2 k^2\ \calk_2-\frac{9 \tf_2^2 f_1^2 }{h (1-f_1^2)^{1/2} f_2}\ \calk_2''\\
&+\frac{9 \tf_2  f_1}{g_s (1-f_1^2)^{3/2} f_2 h^2}\biggl(h \tf_2 f_1^3 g_s'-\tf_2 f_1^3 g_s h'+f_1^3 g_s h 
\tf_2'-\tf_2 f_1^2 g_s h f_1'-f_1 g_s h \tf_2'\\
&-h \tf_2 f_1 g_s'+\tf_2 f_1 g_s h'-\tf_2 g_s h f_1'
\biggr)
\ \calk_2'
+ \frac{9(2 f_3^2 g_s h \calk_2-K \calk_1+4 f_3 g_s h F) f_1^2}{2h^2 f_2 (1-f_1^2)^{1/2} g_s f_3^2}\,.
\end{split}
\eqlabel{k2equa}
\end{equation}

To make use of the results in \cite{hyd3,kt3} we use a radial coordinate $x$ 
as in \eqref{xks}. The physical fluctuations described by \eqref{fequa}-\eqref{k2equa}
must satisfy an incoming wave boundary condition at the horizon of the KT BH, 
and be normalizable at the asymptotic $x\to 0_+$ boundary.
Introducing 
\begin{equation}
\ww=\frac{\w}{2\pi T}\,,\qquad \kk=\frac{k}{2\pi T}\,.
\eqlabel{wq}
\end{equation} 
The former condition implies 
\begin{equation}
F=(1-x)^{-i\ww}\ \hat{F}\,,\qquad \calk_1=(1-x)^{-i\ww}\ \hat{\calk}_1\,,\qquad \calk_2
=(1-x)^{-i\ww}\ \hat{\calk}_2\,,\qquad 
\eqlabel{incoming}
\end{equation}
with $\{\hat{F}\,, \hat{\calk}_1\,, \hat{\calk}_2\}$ being regular at the horizon, \ie 
as $x\to 1_-$. Given \eqref{incoming}, the equations of motion 
for $\{\hat{F}\,, \hat{\calk}_1\,, \hat{\calk}_2\}$ are complex --- they become real 
once we introduce 
\begin{equation}
\ww=-i\ \WW\,,\qquad \Im(\Omega)=0\,.
\eqlabel{WW}
\end{equation}

Using the asymptotic expansion for the KT BH developed in 
\cite{kt3}\footnote{As explained in \cite{kt3} we can set in numerical analysis $a_0=1$.}, the normalizability condition 
for $\{\hat{F}\,, \hat{\calk}_1\,, \hat{\calk}_2\}$
at the $x\to 0_+$ boundary translates into the following asymptotic solution\footnote{For 
numerical analysis we developed all expansions to order $\calo(x^{11/4}\ln^5 x)$ inclusive.}  
\begin{equation}
\begin{split}
&\hat{F}=x^{3/4} F_{3,0}+\biggl(\frac{\sqrt{2}}{32}\biggl((2\pi T\WW)^2+(2\pi T\kk)^2\biggr) 
\left(F_{3,0} k_s+5 F_{3,0}-\calk_{1,3,0}\right)\\
&-\frac{\sqrt{2}}{32} \biggl((2\pi T\WW)^2+(2\pi T\kk)^2\biggr) F_{3,0}\ \ln x\biggr) x^{5/4}
+\biggl(F_{7,0}+\biggl(\frac 67 F_{3,0} a_{2,0}\\
&+(2\pi T\WW)^4\left(-\frac{89}{18432}  F_{3,0}
-\frac{23}{18432}  F_{3,0} k_s
+\frac{13}{18432} \calk_{1,3,0}\right)
\\
&-\left(\frac{23}{9216} F_{3,0} k_s-\frac{13}{9216} \calk_{1,3,0}+\frac{89}{9216}
 F_{3,0}\right) (2\pi T\WW)^2 (2\pi T\kk)^2\\
&+\left(-\frac{23}{18432} F_{3,0} k_s+\frac{13}{18432} \calk_{1,3,0}-\frac{89}{18432}
 F_{3,0}\right) (2\pi T \kk)^4\biggr) \
\ln x\\
&+ \left(\frac{1}{1024} (2\pi T\WW)^2 (2\pi T \kk)^2 F_{3,0}+\frac{1}{2048} (2\pi T\WW)^4 
F_{3,0}+\frac{1}{2048} (2\pi T \kk)^4 F_{3,0}\right)\ \ln^2 x\biggr) x^{7/4}\\
&+\calo\left(x^{9/4}\ln^3 x\right)\,,
\end{split}
\eqlabel{uvf}
\end{equation}
\begin{equation}
\begin{split}
\hat{\calk}_1=x^{3/4} \biggl(\calk_{1,3,0}+\frac 12 F_{3,0}\ \ln x\biggr)+\calo\left(x^{5/4}\ln^2 x\right)\,,
\end{split}
\eqlabel{uvk1}
\end{equation}
\begin{equation}
\begin{split}
\hat{\calk}_2=x^{3/4} \biggl(\frac 32 \calk_{1,3,0}-F_{3,0}+\frac 34 F_{3,0}\ \ln x
\biggr)+\calo\left(x^{5/4}\ln^2 x\right)\,,
\end{split}
\eqlabel{uvk2}
\end{equation}
where we presented the expansions only to leading order in the normalizable 
UV coefficients
\begin{equation}
\biggl\{F_{3,0}\,, F_{7,0}\,, \calk_{1,3,0}\biggr\}\,.
\eqlabel{normalizableuv} 
\end{equation}
The independent UV normalizable coefficients \eqref{normalizableuv} imply that the 
spontaneous $\c$SB in cascading plasma is associated with the development of the 
expectation values of the two dimension-3 operators  --- the gaugino bilinears of the two 
gauge groups ---
\begin{equation}
\calo_3^1\ \propto\ F_{3,0}\,,\qquad  \calo_3^2\ \propto\ \calk_{1,3,0} \,,
\eqlabel{gaugiexp}
\end{equation} 
and a certain dimension-7 operator\footnote{It is difficult to 
identify precisely what is this operator on the gauge theory side
of the correspondence. We expect that this operator is not chiral, see section 4.2.2 for more details.} 
\begin{equation}
\calo_7\ \propto F_{7,0}\,.
\eqlabel{dim7vev}
\end{equation}

Since the equations of motion \eqref{fequa}-\eqref{k2equa} are homogeneous,
without the loss of generality we can set $\hat{F}(1)=1$. The IR, \ie as $y\equiv (1-x)
\to 0_+$, 
asymptotic expansion then takes form\footnote{For numerical analysis we developed 
all expansions to order $\calo(y^6)$ inclusive.}  
\begin{equation}
\hat{F}=1+\calo(y^2)\,,\qquad \hat{\calk}_1=\calk_{1,0}^h+\calo(y^2)
\,,\qquad \hat{\calk}_2=\calk_{2,0}^h+\calo(y^2)\,,
\eqlabel{irasymptotic}
\end{equation}
where we presented the expansions only to leading order in the normalizable 
IR coefficients
\begin{equation}
\biggl\{\calk_{1,0}^h\,, \calk_{2,0}^h\biggr\}\,.
\eqlabel{normalizableir} 
\end{equation}

Notice that altogether we have 5 adjustable parameters: \eqref{normalizableuv}
and \eqref{normalizableir}, in order to solve a boundary value problem for a 
system of 3 second-order differential equations for 
$\{\hat{F}\,, \hat{\calk}_1\,, \hat{\calk}_2\}$.
As a result, a solution produces a dispersion relation for the $\c$SB
quasinormal modes:
\begin{equation}
\Omega=\Omega(\kk^2)\,.
\eqlabel{disp}
\end{equation}
The quasinormal modes signal an instability in plasma provided 
\begin{equation}
\Im(\ww)>0\ \Leftrightarrow\ \Omega<0\,,\qquad {\rm provided}\qquad \Im(\kk)=0\,.
\eqlabel{instablity}
\end{equation}

\begin{figure}[t]
\begin{center}
\psfrag{lnTL}{{$\ln \frac{T}{\Lambda}$}}
\psfrag{qq}{{$\kk^2$}}
\includegraphics[width=4in]{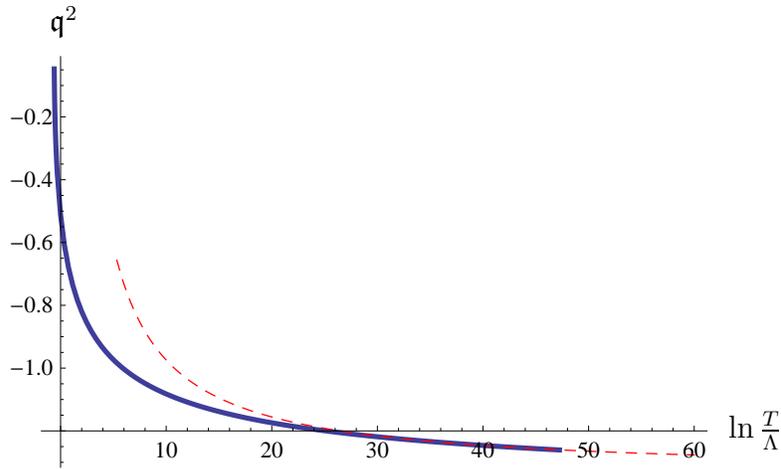}
\end{center}
  \caption{(Colour online) Dispersion relation of the $\c$SB quasinormal modes 
of the Klebanov-Tseytlin black hole as a function of $\ln \frac{T}{\Lambda}$
at high temperature. The solid blue line represent the dispersion relation of the $\c$SB
fluctuations with $(i\ww=0,\kk^2)$.
The red dashed line is a fit \eqref{fitred} to the data.
 } \label{figure1}
\end{figure}

\begin{figure}[t]
\begin{center}
\psfrag{TL}{{$\frac{T}{\Lambda}$}}
\psfrag{qq}{{$\kk^2$}}
\includegraphics[width=3in]{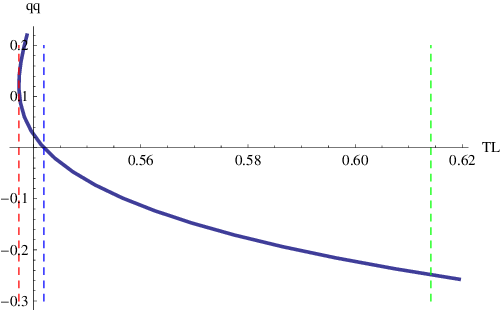}
\includegraphics[width=3in]{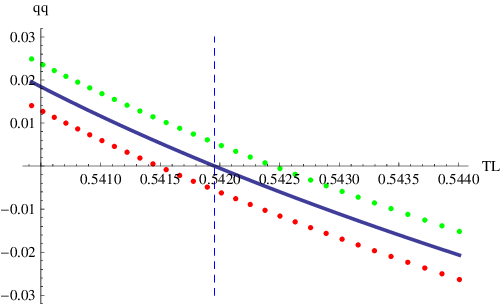}
\end{center}
  \caption{(Colour online) Dispersion relation of the $\c$SB quasinormal modes 
of the Klebanov-Tseytlin black hole as a function of $\frac{T}{\Lambda}$.
The solid blue lines represent the dispersion relation of the $\c$SB
fluctuations at the threshold of instability: $(\ww=0,\kk^2)$.
The blue dashed vertical lines represent the onset of instability: $T=T_{\c\rm{SB}}$,
such that $(i\ww=0,\kk^2=0)$. The vertical dashed green and red lines 
indicate $T=T_c$ and $T=T_u$ correspondingly. The green dots indicate 
quasinormal modes with $(i\ww=0.01, \kk^2)$ as a function of $\frac{T}{\Lambda}$.
 The red dots indicate 
quasinormal modes with $(i\ww=-0.01, \kk^2)$ as a function of $\frac{T}{\Lambda}$.
 } \label{figure2}
\end{figure}

The results of the analysis of the dispersion relation of $\c$SB  
quasinormal modes are presented in Figures~\ref{figure1}-\ref{figure2}.
In principle, we expect discrete branches of the quasinormal modes
distinguished by the number of nodes in radial profiles 
$\{\hat{F}\,,\hat{\calk}_1\,,\hat{\calk}_2\}$. In what follows we consider 
only the lowest quasinormal mode, which has monotonic radial profiles. 

In Figure~\ref{figure1} we study the dispersion relation of the $\c$SB 
quasinormal modes at high temperatures. Here, we expect the KT BH to 
be stable with respect to such fluctuations. Indeed we find that 
that fluctuations with $(\WW=0,\kk^2)$ (solid blue line) have $\kk^2<0$  --- as a result,
they are massive. The red dashed line 
\begin{equation}
\kk^2\bigg|_{\rm{red,dashed}}\ =\ -1.33(7)+3.62(9)\ \ln^{-1}\frac{T}{\Lambda}
+\calo\left(\ln^{-2}\frac{T}{\Lambda}\right) \,,
\eqlabel{fitred}
\end{equation}
represents the best fit to (the high-temperature tail of)  the data. 
Notice that in the limit $T\gg \Lambda$ the cascading theory approaches a 
conformal theory with temperature being the only relevant scale, thus, 
in agreement with \eqref{fitred}, $\kk^2$
must approach a constant in this limit.

Figure~\ref{figure2} presents results for the $\c$SB modes dispersion relation
at low temperatures. 
The solid blue line on the left plot represents the dispersion relation of the 
$\c$SB fluctuations at the threshold of instability, \ie we have $\WW=0$ with $\kk^2\ne 0$. 
Notice that as long as $T> T_{\c\rm{SB}}$ (represented by a vertical blue dashed line)
$\kk^2<0$ for these modes, which makes them massive. As a result, translationary 
invariant KT horizons are stable against chiral symmetry breaking 
fluctuations all the way down to $T_{\c\rm{SB}}$. $T_{\c\rm{SB}}$ is lower than 
the temperature $T_c$ of the confinement/deconfinement phase transition in the cascading plasma 
(represented by a vertical green dashed line), but is above the temperature $T_u$  (represented by a vertical red dashed line)
of the hydrodynamic instability in the deconfined cascading plasma. At  temperatures 
$T<T_{\c\rm{SB}}$ the $\c\rm{SB}$ fluctuations have $\kk^2>0$ --- they are tachyonic. 
The right plot on Figure~\ref{figure2} presents detailed dispersion relation in the 
vicinity of $\c$SB instability. Here, again, the solid blue line represents the 
dispersion relation at the threshold of instability $(i\ww=0,\kk^2)$; 
the vertical dashed blue line defines the temperature of the $\c$SB in the deconfined 
cascading plasma:
\begin{equation}
\frac{T}{\Lambda}\bigg|_{\c\rm{SB}}=0.54195(5)\,.
\eqlabel{tcsb}
\end{equation}
At a given temperature, quasinormal modes with $\kk^2$ below the 
momenta of the modes at the threshold of instability (blue line) 
are expected to have $i\ww<0$ (indicating a genuine tachyonic instability),
while modes with $\kk^2$ above the momenta of the modes at the threshold of instability 
are expected to have $i\ww>0$ (indicating  stable excitations). This is precisely what we 
find:  the red dots on the right plot have $i\ww=-0.01$ and the green dots 
indicate quasinormal modes with $i\ww=0.01$.

\section{Homogeneous and isotropic  end point of  chiral tachyon condensation }
In section 3  we showed that translationary invariant KT horizons describing 
the equilibrium states of the deconfined cascading plasma at strong coupling become unstable 
with respect to the chiral symmetry breaking at $T<T_{\c\rm{SB}}$. In this 
section we ask whether the endpoint of the $\c$SB tachyon condensation in deconfined 
cascading plasma realizes a homogeneous and isotropic equilibrium state. 
In other words, we attempt to 
construct a Klebanov-Strassler black hole (black brane) solution  
with spontaneously broken chiral symmetry. 
We present (a technical) evidence that such solution does not exist. 
A physical argument for the absence of homogeneous and isotropic deconfined 
equilibrium states in the cascading plasma with spontaneously broken chiral symmetry 
is presented in section 5\footnote{A reader interested in this physical argument 
would still need the results of sections 4.1-4.4.}.
   
This section is organized as follows. We introduce parametrization of the 
(generically mass-deformed\footnote{See below.}) KS BH 
closely resembling that of KT BH of \cite{kt3} in section 4.1. The general UV 
asymptotics obtained from solving equations of motion \eqref{ks1}-\eqref{ksc}
are given in section 4.2. We further identify normalizable and non-normalizable 
parameters of the UV asymptotics. In particular, we identify two independent 
mass parameters dual to the gaugino masses in the cascading gauge theory.  
Thus, a generic solution of \eqref{ks1}-\eqref{ksc} with a regular Schwarzschild 
horizon represents a homogeneous and isotropic  
equilibrium state  in {\it mass-deformed} cascading gauge theory plasma. Because such masses 
are introduced for the fermions of  $\caln=1$ vector multiplet of cascading gauge theory,
they explicitly break both the supersymmetry (at zero temperature) and chiral symmetry.   
The general IR asymptotics of \eqref{ks1}-\eqref{ksc} guaranteeing a regular  Schwarzschild 
horizon are presented in section 4.3. In section 4.4 we perform the general parameter 
counting for the numerical analysis. Finally, in section 4.5, 
insisting on vanishing mass parameters for the gauginos of the cascading gauge theory,  
we {\it deform} KT BH solution in the direction of the $\c$SB tachyons. 
For zero mass parameters there is always a solution to  \eqref{ks1}-\eqref{ksc}: 
namely, the KT BH with vanishing chiral condensates.  
One would expect that a sufficiently large deformation 
in the {\it tachyonic direction}  would lead to a new solution with 
non-vanishing chiral condensates. As alluded to earlier, we do not find 
such new solution.

\subsection{Parametrization of the KS BH}

Let's fix the radial coordinate as in \eqref{xks}
\begin{equation}
x=1-f_1(r)\,,
\eqlabel{gauge}
\end{equation}
and introduce new functions $\{K_1,\ K_2,\ K_3,\ f_a,\ f_b,\ f_c,\ h,\ g\}$ as follows:
\begin{equation}
\begin{split}
&h_1(x)=\frac 1P \left(\frac{1}{12}\ K_1(x)-36 \Om_0\right)\,,\qquad h_2(x)=\frac{P}{18}\ K_2(x)\,,\\
&h_3(x)=\frac 1P \left(\frac{1}{12}\ K_3(x)-36 \Om_0\right)\,,\qquad g_s(x)=g(x)\,,\\
&\w_1(x)=\frac{\sqrt{f_c(x)}}{3(2x-x^2)^{1/4}}\,,\qquad \w_2(x)=\frac{\sqrt{f_a(x)}}{\sqrt{6}(2x-x^2)^{1/4}}\,,
\qquad \w_3(x)=\frac{\sqrt{f_b(x)}}{\sqrt{6}(2x-x^2)^{1/4}}\,,\\
&H(x)=(2x-x^2)\ h(x)\,.
\end{split}
\eqlabel{deff}
\end{equation}
We can then rewrite equations \eqref{ks1}-\eqref{ksc} and find that $\Om_0$ disappears.

\subsection{UV asymptotics}
The UV asymptotic corresponds to $x\to 0_+$. We find:
\begin{equation}
K_1=4 h_0a_0^2-\frac 12 P^2 g_0-\frac12 P^2 g_0\ \ln x+\sum_{n=1}^\infty\sum_k\ k_{1nk}\ x^{n/4}\ \ln^k x\,,
\eqlabel{uv1}
\end{equation}
\begin{equation}
K_2=1+\sum_{n=1}^\infty\sum_k\ k_{2nk}\ x^{n/4}\ \ln^k x\,,
\eqlabel{uv2}
\end{equation}
\begin{equation}
K_3=4 h_0a_0^2-\frac 12 P^2 g_0-\frac12 P^2 g_0\ \ln x+\sum_{n=1}^\infty\sum_k\ k_{3nk}\ x^{n/4}\ \ln^k x\,,
\eqlabel{uv3}
\end{equation}
\begin{equation}
f_a=a_0\biggl(1+\sum_{n=1}^\infty\sum_k\ f_{ank}\ x^{n/4}\ \ln^k x\biggr)\,,
\eqlabel{uv4}
\end{equation}
\begin{equation}
f_b=a_0\biggl(1+\sum_{n=1}^\infty\sum_k\ f_{bnk}\ x^{n/4}\ \ln^k x\biggr)\,,
\eqlabel{uv5}
\end{equation}
\begin{equation}
f_c=a_0\biggl(1+\sum_{n=2}^\infty\sum_k\ f_{cnk}\ x^{n/4}\ \ln^k x\biggr)\,,
\eqlabel{uv6}
\end{equation}
\begin{equation}
h=h_0-\frac  {P^2 g_0}{8a_0^2}\ \ln x+\sum_{n=2}^\infty\sum_k\ h_{nk}\ x^{n/4}\ \ln^k x\,,
\eqlabel{uv7}
\end{equation}
\begin{equation}
g=g_0\left[\ 1+\sum_{n=2}^\infty\sum_k\ g_{nk}\ x^{n/2}\ \ln^k x \ \right]\,.
\eqlabel{uv8}
\end{equation}

We developed UV expansion to order $n=8$ inclusive. 
The expansion depends on 5 microscopic parameters
\begin{equation}
\{g_0\,,a_0\,,\ h_0\,, k_{110}\,, f_{a10}\}\,,
\eqlabel{par}
\end{equation}
where $g_0$ is related to the dimensionless parameter of the cascading gauge theory, and as we explain below,
the four independent combinations of the other parameters are related to the temperature, the dynamical scale of the cascading 
gauge theory, and the two mass parameters (the couplings of the two dimension-3 operators that explicitly break the chiral symmetry 
of the cascading theory).  
Besides \eqref{par}, the expansions \eqref{uv1}-\eqref{uv8} are characterized by 7 vev's:
\nxt those of dimension-3 operators:
\begin{equation}
\{f_{a30}\,,\ k_{230} \}\,,
\eqlabel{vev3}
\end{equation}
\nxt those of dimension-4 operators:
\begin{equation}
\{f_{a40}\,,\ g_{40} \}\,,
\eqlabel{vev4}
\end{equation}
\nxt that of a dimension-6 operator:
\begin{equation}
\{f_{a60} \}\,,
\eqlabel{vev6}
\end{equation}
\nxt that of a dimension-7 operator:
\begin{equation}
\{f_{a70} \}\,,
\eqlabel{vev7}
\end{equation}
\nxt and finally, that of a dimension-8 operators:
\begin{equation}
\{f_{a80} \}\,.
\eqlabel{vev8}
\end{equation}
Note that characterization in \eqref{vev3}-\eqref{vev8} is suggestive only --- typically,
a combination of operators is mapped to a given normalizable mode of a dual gravitational field. 
In what follows we will not need the precise map between the operator vevs and corresponding 
normalizable coefficients of the dual geometry\footnote{The precise map for dimension-4 
operators of chirally symmetric states of the cascading gauge theory is discussed in \cite{aby}.}.

\subsubsection{Mass-deformed KS solution at $T=0$}
In order to understand the physical meaning of the microscopic parameters  \eqref{par}
we develop the asymptotic solution of \eqref{ks1}-\eqref{ksc} in the limit $T=0$,  {\it i.e,} for 
\begin{equation}
f_1\equiv 1\,.
\eqlabel{ksm0}
\end{equation}
Here, 
the radial coordinate \eqref{gauge} is undefined, so instead we fix the gauge as 
\begin{equation}
\tf_2=\frac r3\,.
\eqlabel{f2ksm}
\end{equation}
Similarly to \eqref{deff}, we introduce 
\begin{equation}
\begin{split}
&h_1(r)=\frac 1P \left(\frac{1}{12}\ \tK_1(r)-36 \Om_0\right)\,,\qquad h_2(r)=\frac{P}{18}\ \tK_2(r)\,,\\
&h_3(r)=\frac 1P \left(\frac{1}{12}\ \tK_3(r)-36 \Om_0\right)\,,\qquad g_s(r)=\tg(r)\,,\\
&\w_1(r)=\frac{\sqrt{\tf_c(r)}}{3r}\,,\qquad \w_2(r)=\frac{\sqrt{\tf_a(x)}}{\sqrt{6}r}\,,
\qquad \w_3(r)=\frac{\sqrt{\tf_b(r)}}{\sqrt{6}r}\,,\\
&H(x)=r^4 \th(r)\,,
\end{split}
\eqlabel{deff0}
\end{equation}
and find that $\Om_0$ disappears from \eqref{ks1}-\eqref{ksc}.

The UV asymptotic corresponds to $r\to 0_+$. We find:
\begin{equation}
\tK_1=4 \th_0 -\frac 12 P^2 \tg_0-2 P^2 \tg_0\ \ln r+\sum_{n=1}^\infty\sum_k\ \tk_{1nk}\ r^n\ \ln^k r\,,
\eqlabel{uv10}
\end{equation}
\begin{equation}
\tK_2=1+\sum_{n=1}^\infty\sum_k\ \tk_{2nk}\ r^n\ \ln^k r\,,
\eqlabel{uv20}
\end{equation}
\begin{equation}
\tK_3=4 \th_0-\frac 12 P^2 \tg_0-2 P^2 \tg_0\ \ln r+\sum_{n=1}^\infty\sum_k\ \tk_{3nk}\ r^n\ \ln^k r\,,
\eqlabel{uv30}
\end{equation}
\begin{equation}
\tf_a=1+\sum_{n=1}^\infty\sum_k\ \tf_{ank}\ r^n\ \ln^k r\,,
\eqlabel{uv40}
\end{equation}
\begin{equation}
\tf_b=1+\sum_{n=1}^\infty\sum_k\ \tf_{bnk}\ r^n\ \ln^k r\,,
\eqlabel{uv50}
\end{equation}
\begin{equation}
\tf_c=1+\sum_{n=2}^\infty\sum_k\ \tf_{cnk}\ r^n\ \ln^k r\,,
\eqlabel{uv60}
\end{equation}
\begin{equation}
\th=\th_0-\frac 12 P^2 \tg_0\ \ln r+\sum_{n=2}^\infty\sum_k\ \th_{nk}\ r^n\ \ln^k r\,,
\eqlabel{uv70}
\end{equation}
\begin{equation}
\tg=\tg_0\left[\ 1+\sum_{n=2}^\infty\sum_k\ \tg_{nk}\ r^n\ \ln^k r \ \right]\,.
\eqlabel{uv80}
\end{equation}
Here, the expansion depends on 4 microscopic parameters: 
\nxt the asymptotic string coupling $\tg_0$
\nxt the two mass deformation parameters $\tk_{110}, \tf_{a10}$
\nxt and the parameter $\th_0$, related to the strong coupling scale of the cascading gauge theory, 
see \cite{aby}. 
Given a set $\{\tg_0, \tk_{110}, \tf_{a10}, \th_0\}$ the UV solution is uniquely determined by 
the condensates of various relevant and irrelevant operators, in analogy with \eqref{vev3}-\eqref{vev8}:
\begin{equation}
\{\tf_{a30}\,,\tk_{230}\,,\tf_{a40}\,,\tg_{40}\,,\tf_{a60}\,,\tf_{a70}\,,\tf_{a80}\}\,.
\eqlabel{masscond0}
\end{equation}
We emphasize that while turning on the finite temperature the four microscopic parameters of the theory 
must be kept fixed; on the other hand, the condensates \eqref{masscond0} will develop a nontrivial temperature dependence.

We would like to match \eqref{uv10}-\eqref{uv80} with the asymptotic finite-temperature solution \eqref{uv1}-\eqref{uv8}. 
We require that as $r\to 0$ (and correspondingly $x\to 0$) all the corresponding warp factors in the metric 
should agree to leading order, {\it i.e.},  
\begin{equation}
\begin{split}
&\lim_{\{r,x\}\to0} \frac{r^4 \th(r)}{(2x-x^2)h(x)}=1\,,\qquad 
\lim_{\{r,x\}\to 0}\frac{\th(r)^{1/2}\tf_{a,b,c}(r)}{h(x)^{1/2}f_{a,b,c}(x)}=1\,,\\
&\lim_{\{r,x\}\to0} \frac{\tK_{1,2,3}(r)}{K_{1,2,3}(x)}=1\,,\qquad \lim_{\{r,x\}\to0}\frac{\tg(r)}{g(x)}=1\,.
\end{split}
\eqlabel{matching}
\end{equation}
This matching uniquely identifies:
\begin{equation}
\begin{split}
&x=\frac 12 a_0^2 r^4+{\rm higher\ order}\,,\qquad g_0=\tg_0\,,\qquad h_0 a_0^2=\th_0+\frac 18 P^2 \tg_0\ln( \frac{a_0^2}{2})\,,\\
&k_{110}=\frac{2^{1/4}}{a_0^{1/2}}\ \tk_{110}\,,\qquad  f_{a10}=\frac{2^{1/4}}{a_0^{1/2}}\ \tf_{a10}+\frac{2^{1/4}}{a_0^{1/2}}\ 
\frac{2 \tk_{110}+P^2\tg_0 \tf_{a10}}{3 P^2 \tg_0+8\th_0}\ \ln(\frac{a_0^2}{2})\,.
\end{split}
\eqlabel{resmatch}
\end{equation}
From \eqref{resmatch} we see that keeping the microscopic parameters $\{\tg_0,\th_0,\tk_{110},\tf_{a10}\}$ of the mass-deformed cascading gauge theory fixed
requires that at finite temperature the following four (corresponding) combinations of  \eqref{par} must be kept fixed 
\begin{equation}
\begin{split}
\left\{g_0\,,  \biggl[ h_0\ a_0^2-\frac 18P^2 g_0 \ln(\frac{a_0^2}{2})\biggr]\,, \biggl[k_{110}\ a_0^{1/2}\biggr]\,, \biggl[f_{a10}-
\frac{2k_{110}+P^2 g_0 f_{a10}}{3 P^2 g_0+8 h_0 a_0^2}\ \ln(\frac{a_0^2}{2})\biggr]\ a_0^{1/2}
\right\}\,.
\end{split}
\eqlabel{fixed}
\end{equation}

\subsubsection{Comparison with the extremal KS solution}
To compare the asymptotic expansion \eqref{uv10}-\eqref{uv80} with the extremal KS solution \eqref{ksks} we need to 
turn off both the mass deformation parameters
\begin{equation}
\tk_{110}=\tf_{a10}=0\,.
\eqlabel{masszero}
\end{equation}
We further denote the radial coordinate of the extremal KS solution \eqref{ksks}, \eqref{kk} as $r_{KS}$.
Then with 
\begin{equation}
\begin{split}
e^{-r_{KS}}\equiv& \frac 38\,\sqrt {6}{\epsilon}^{2}{r}^{3} \left( 1+\calo(r^{12}\ln r)\right)\,,
\end{split}
\eqlabel{rksr}
\end{equation}
we can identify asymptotic expansions \eqref{uv10}-\eqref{uv80} with  \eqref{ksks}, \eqref{kk} to order $\calo(r^{9})$ provided:
\begin{equation}
\begin{split}
&\th_0=\frac{1}{24}\,P^2 { g_0} \left( 10\,\ln  \left( 2 \right) -6\,\ln 
 \left( 3 \right) -4\,\ln  \left( {\epsilon}^{2} \right) -1 \right)\,,\qquad \tg_0=g_0 
\,,\\
&\tf_{a40}=\tg_{40}=\tf_{a70}=\tf_{a80}=0\,,\\
&\tf_{a30}=\frac 34 \sqrt{6} \e^2\,,\qquad \tk_{230}=-\frac 38\,\sqrt {6}{\epsilon}^{2} \left( 5\,\ln  \left( 2 \right) -3\,\ln 
 \left( 3 \right) -2\,\ln  \left( {\epsilon}^{2} \right)  \right)\,,\\
&\tf_{a60}=-{\frac {9}{16}}\,{\epsilon}^{4} \left( -3+5\,\ln  \left( 2 \right) -3
\,\ln  \left( 3 \right) -2\,\ln  \left( {\epsilon}^{2} \right) 
 \right) \,.
\end{split}
\eqlabel{kscond}
\end{equation}
Notice that in a supersymmetric ground state of the cascading gauge theory (a Klebanov-Strassler solution)
the expectation value of dimension-7 operator $\calo_7\propto \tf_{a70}$ vanishes. It is thus suggestive 
that this operator is not chiral.   

A further  rescaling of the radial coordinate \eqref{f2ksm}
\begin{equation}
r\to \hat{r}=r \e^{2/3}\,,
\eqlabel{res1}
\end{equation}
would modify the asymptotic UV-parameters 
\begin{equation}
\begin{split}
\{\tg_0,\th_0,\tk_{230},\tf_{a30},&\tf_{a40},\tf_{a60},\tf_{a70},\tf_{a80},\tg_{40}\} \,,\\
&\Downarrow\\
\{\hat{g}_0,\hat{h}_0,\hat{k}_{230},\hat{f}_{a30},&\hat{f}_{a40},
\hat{f}_{a60},\hat{f}_{a70},\hat{f}_{a80},\hat{g}_{40}\}\,,
\end{split}
\eqlabel{res2}
\end{equation}
in such a way that UV-parameters of the KS solution \eqref{kscond} would take a particularly simple form (note that any reference to the 
KS scale parameter $\e$, as in \eqref{ksks}, disappears): 
\begin{equation}
\begin{split}
&e^{-r_{KS}}=\frac 38\,\sqrt {6}{\hat{r}}^{3} \left( 1+\calo(\hat{r}^{12}\ln(\hat{r}))\right)\,,
\\
&\hat{h}_0=\frac{1}{24}\,P^2 { g_0} \left( 10\,\ln  \left( 2 \right) -6\,\ln 
 \left( 3 \right) -1 \right)\,,\qquad \hat{g}_0=g_0 \,,
\\
&\hat{f}_{a40}=\hat{g}_{40}=\hat{f}_{a70}=\hat{f}_{a80}=0\,,\\
&\hat{f}_{a30}=\frac 34 \sqrt{6}\,,\qquad \hat{k}_{230}=-{\frac {15}{8}}\,\sqrt {6}\ln  \left( 2 \right) +{\frac {9}{8}}\,
\sqrt {6}\ln  \left( 3 \right) \,,
\\
&\hat{f}_{a60}={\frac {27}{16}}-{\frac {45}{16}}\,\ln  \left( 2 \right) +{\frac {27}{
16}}\,\ln  \left( 3 \right) \,.
\end{split}
\eqlabel{kscond1}
\end{equation}

\subsubsection{Comparison with KT BH}
It is straightforward to relate the UV parameters of the KS BH \eqref{par}-\eqref{vev8} 
with that of the KT BH \cite{kt3}.
First, we need to set 
\begin{equation}
k_{110}=f_{a10}=f_{a30}=k_{230}=f_{a70}=0\,.
\eqlabel{ktbh1}
\end{equation} 
The radial coordinate \eqref{gauge} is identical to the one used in \cite{kt3}.
Thus, relating the asymptotic expansions \eqref{uv1}-\eqref{uv8} with the corresponding expressions 
in \cite{kt3} we find:
\begin{equation} 
\begin{split}
&h_0=h_{0,0}\,,
\end{split}
\eqlabel{ktbh2}
\end{equation} 
\begin{equation}
\begin{split}
&g_{40}=\frac{g_{2,0}}{g_0}\,,
\end{split}
\eqlabel{ktbh4}
\end{equation} 
\begin{equation}
\begin{split}
&f_{a40}=-\frac 17\,\frac{ a_{2,0}}{a_0}\,,
\end{split}
\eqlabel{ktbh5}
\end{equation} 
\begin{equation}
\begin{split}
&f_{a60}=-\frac 14\ \frac{a_{3,0}}{a_0}\,,
\end{split}
\eqlabel{ktbh6}
\end{equation} 
\begin{equation}
\begin{split}
&f_{a80}=\frac 1\dd\biggl\{
P^2 g_0\biggl(\frac{6366}{245} \left(\frac{{a}_{2,0}}{a_0}\right)^2-6 \left(\frac{g_{2,0}}{g_0}\right)^2+6+\frac{74}{7}\ \frac{a_{2,0}}{a_0}
-41\ \frac{a_{4,0}}{a_0}\biggr)+h_{0,0}\biggl(\frac{480}{7}\ \frac{a_{2,0}}{a_0}\\
&+\frac{126144}{245}\ \left(\frac{a_{2,0}}{a_0}\right)^2-120\ \frac{a_{4,0}}{a_0}-\frac{576}{7}\ \frac{a_{2,0}}{a_0}\ \frac{g_{2,0}}{g_0}\biggr)
+\frac{13824}{49}\ \left(\frac{a_{2,0}}{a_0}\right)^2\ \frac{h_{0,0}^2}{P^2 g_0}\biggr\}\,,
\end{split}
\eqlabel{ktbh7}
\end{equation} 
where 
\begin{equation}
\dd=139 P^2 g_0-120 h_{0,0}a_0^2\,.
\eqlabel{ktbh8}
\end{equation}

\subsection{IR asymptotics}
Introducing $y=1-x$, the regular horizon $y\to 0_+$ asymptotics of 
$$\{K_1,\ K_2,\ K_3,\ f_a,\ f_b,\ f_c,\ h,\ g\}\,,$$ (defined as in \eqref{deff}) take form:
\begin{equation}
\begin{split}
&K_i=\sum_{n=0}^\infty k_{ihn}\ y^{2n}\,,\qquad i=1,2,3\,,\\
&f_{\alpha}=a_0\ \sum_{n=0}^\infty f_{\a hn}\ y^{2n}\,,\qquad \a=a,b,c\,,\\
&h=\sum_{n=0}^\infty h_{hn}\ y^{2n}\,,\qquad g=g_0\ \sum_{n=0}^\infty g_{hn}\ y^{2n}\,.
\end{split}
\eqlabel{ir1}
\end{equation}
We developed IR expansion to order $n=1$ inclusive. 
Here, the expansion is characterized by 9 parameters:
\begin{equation}
\{k_{1h0}\,, k_{2h0}\,, k_{3h0}\,,f_{ah0}\,, f_{ah1}\,, f_{bh0}\,, f_{ch0}\,, h_{h0}\,, g_{h0}\}\,.
\eqlabel{ircond}
\end{equation}

\subsubsection{Comparison with KT BH}
By matching the near-horizon asymptotic expansions \eqref{ir1} with the corresponding ones in \cite{kt3}, 
we can relate 
\eqref{ircond} to those of the KT BH solution. We find:
\begin{equation}
\begin{split}
&h_{h0}=h^h_0\,,\qquad g_{h0}=\frac{g^h_0}{g_0}\,,\qquad k_{1h0}=k_{3h0}=k^{h}_0\,,\qquad k_{2h0}=1\,,\\
&f_{ah0}=f_{bh0}=\frac{b^h_0}{a_0}\,,\qquad f_{ch0}=\frac{a^h_0}{a_0}\,,\\
&f_{ah1}=\frac{1}{a_0\dd_h}\biggl[\left(2 h^h_0a^h_0\ (3 a^h_0+2 a^h_1)-\frac 12 P^2 g^h_0\right)b^h_0
+6h^h_0\ (a^h_0+2 a^h_1)\ \left(b^h_0\right)^2\biggr]\,,
\end{split}
\eqlabel{ksircond}
\end{equation}
where as in \cite{kt3}
\begin{equation}
\dd_h\equiv 8 h^h_0\ \left(a^h_0\right)^2-P^2 g^h_0 \,.
\eqlabel{defdh}
\end{equation}

\subsection{Parameter counting and the numerical procedure}
In this section we would like to further understand the physical meaning of the microscopic parameters 
\eqref{par}. As we mentioned,
$g_0 P^2$ is the dimensionless parameter of the cascading theory
(which must be large for the gravity approximation to be valid),
while $a_0\,, h_{0}\,, k_{110}$ and $f_{a10}$ are related to the strong coupling scale of cascading
theory, the two mass parameters, and to the temperature.

Much like in \cite{kt3}, it can be shown that 
\begin{equation}
a_0^2=4\pi G_5\ sT\,,
\eqlabel{a0mean}
\end{equation}
where $T$ is the temperature of the black hole, $s$ is its entropy density, and the effective five-dimensional 
Newton's constant $G_5$ is given by \eqref{g5deff}.

Following \cite{kt3}, we introduce a new dimensionless coefficient  $k_s$ as 
\begin{equation}
P^2 g_0 k_s=4 h_0 a_0^2 -\frac 12 P^2 g_0\,.
\eqlabel{ksdef}
\end{equation}
The second constraint in \eqref{fixed} then implies that the combination $\left[k_s-\frac 12 \ln (\frac{a_0^2}{2})\right]$
is independent of the temperature. Thus, we can choose it to define the strong coupling scale $\Lambda$ of 
the cascading theory:
\begin{equation}
k_s\equiv \frac 12\ \ln \left(\frac{a_0^2}{\Lambda^4}\right)=\frac 12\ \ln \left(\frac{4\pi G_5 sT}{\Lambda^4}\right)\,.
\eqlabel{defl}
\end{equation}
Using the expressions for the high temperature entropy density of
the theory computed in \cite{kt3}, we see that at high
temperatures $k_s\simeq (1/2) \ln(T^4/\Lambda^4)$, with corrections
scaling as $\ln(\ln(T/\Lambda))$. We will use $k_s$ instead of the
temperature as our basic dimensionless parameter, and use
\eqref{defl} to translate between $k_s$ and $T/\Lambda$.
 
Further introducing $K_{110}$ via the relation
\begin{equation}
k_{110}=K_{110}\ \left(3P^2 g_0+8h_0 a_0^2\right)-\frac 12 P^2 g_0 f_{a10}\,,
\eqlabel{k110def}
\end{equation}
the remaining constraints in \eqref{fixed} are solved with 
\begin{equation}
f_{a10}=\left(\mu_1+4 \mu_2\ k_s\right)\ e^{-k_s/2}\,,\qquad K_{110}=\mu_2\ e^{-k_s/2}\,,
\eqlabel{solfixed}
\end{equation}
where $\mu_i$ are the fixed (reduced) mass-deformation parameters of the cascading 
gauge theory
\begin{equation}
\mu_i\equiv \frac{m_i}{\Lambda}\,,\qquad m_i={\rm constant}\,.
\eqlabel{defmu}
\end{equation}
From \eqref{solfixed} we see that at high temperatures, $T\gg \Lambda$, 
\begin{equation}
f_{a10}\simeq \frac{1}{T}\left({m_1}+8\ {m_2}\ \ln(\frac{T}{\Lambda}) \right)\,,\qquad K_{110}\simeq \frac{m_2}{T}\,.
\end{equation}

Note that our metric ansatz (see \eqref{metricks}, \eqref{deff}) is
invariant under a scaling symmetry taking
\begin{equation} (t,x_1,x_2,x_3) \to \lambda^{-1/2}\ (t,x_1,x_2,x_3)\,,\qquad
h\to \l^{-2}\ h\,,\qquad  f_{a,b,c}\to \l\ f_{a,b,c}\,,
\eqlabel{scaling}
\end{equation}
and leaving all other functions in our solution (as well as the
coordinate $x$) invariant. 
We can now use the scaling symmetry \eqref{scaling} to set 
\begin{equation}
a_0=1\,.
\eqlabel{swta0}
\end{equation}
Recall also that we are solving the theory in the supergravity
approximation, which includes only the leading order terms both in
the $g_s$ expansion and in the curvature ($\alpha'$) expansion. When
we neglect $g_s$ corrections, the action (and the equations of
motion we wrote) does not depend separately on $P^2$ and $g$ but
only on the combination $P^2 g$. We can thus set $g_0=1$, and recall
that whenever we have a factor of $P^2$ we really mean $P^2 g_0$.
Furthermore, when we neglect $\alpha'$ corrections, the action is
multiplied by a constant when we rescale the ten dimensional metric
by a constant factor (and rescale the $p$-forms accordingly), so
that the equations of motion are left invariant; this transformation
acts on our variables as
\begin{equation} \eqlabel{scaling2}
h\to \lambda^{2} h\,,\qquad f_{a,b,c} \to f_{a,b,c}\,,\qquad K_{1,3}\to
\lambda^2 K_{1,3}\,,\qquad K_2\to K_2\,,\qquad g\to g\,,
\end{equation}
and it changes $P$ by $P\to \lambda P$. We can use this
transformation to relate the solutions for different values of $P$
(as long as we are in the supergravity approximation). Thus, we will
perform the numerical analysis for $P=1$, and we can use
\eqref{scaling2} to obtain the solutions for any other value of $P$.

We are now ready to formulate our numerical procedure, and count the parameters of the 
solution:
\nxt We integrate the differential equations along $x$-coordinate
\begin{equation}
0\le x \le 1\,,
\eqlabel{xrange}
\end{equation}
with $x=0$ being the boundary and $x=1$ being the horizon. 
\nxt We use various scaling symmetries discussed above to set
\begin{equation}
P=g_0=a_0=1\,.
\eqlabel{reduce}
\end{equation}
\nxt Altogether we need to integrate 8 functions 
\begin{equation}
\{K_1,K_2,K_3,f_a,f_b,f_c,h,g\}\,,
\eqlabel{functions}
\end{equation}
for a given set of the remaining microscopic parameters 
$\{k_s\,, f_{a10}\,, K_{110}\}$\footnote{We can always use \eqref{defl}, \eqref{solfixed}, and \eqref{defmu}
to convert these parameters into the physical temperature and the masses.
}. 
\nxt The solution is  then determined by 7 UV parameters \eqref{vev3}-\eqref{vev8}, and 9 IR parameters \eqref{ircond}:  
\begin{equation}
\begin{split}
{\rm UV}:&\qquad \{f_{a30}\,, k_{230}\,, f_{a40}\,, g_{40}\,, f_{a60}\,, f_{a70}\,, f_{a80}\}\,,\\
{\rm IR}:&\qquad \{k_{1h0}\,, k_{2h0}\,, f_{3h0}\,, f_{ah0}\,, f_{ah1}\,, f_{bh0}\,, f_{ch0}\,, h_{h0}\,, g_{h0}\}\,.
\end{split}
\eqlabel{paruvir}
\end{equation} 
Overall we have 16 parameters, precisely what is necessary to determine \eqref{functions} from the appropriate second 
order differential equations.

We follow numerical method introduced in \cite{kt3}. In a nutshell, for a fixed set of 
microscopic parameters $\{k_s,f_{a10},K_{110}\}$, we choose a 'trial' set of parameters \eqref{paruvir}
and integrate (a double set of) the equations of motion for \eqref{functions} from the 
UV ($x_{initial}=0.01$)  to $x=0.5$, and from the IR ($y_{initial}=0.01$) to $y=0.5$. 
A solution \eqref{paruvir} of the boundary value problem implies that the mismatch vector
\begin{equation}
\begin{split}
\vec{v}_{mismatch}\equiv& \biggl(K_1^b-K_1^h\,,\ (K_1^b+K_1^h)'\,,\ K_2^b-K_2^h\,,\ (K_2^b+K_2^h)'\,,\ 
K_3^b-K_3^h\,,\ \\
&(K_3^b+K_3^h)'\,,\ f_a^b-f_a^h\,,\ (f_a^b+f_a^h)'\,,\ f_b^b-f_b^h\,,\ (f_b^b+f_b^h)'\,,\ 
f_c^b-f_c^h\,,\ \\
&(f_c^b+f_c^h)'\,,\ h^b-h^h\,,\ (h^b+h^h)'\,,\ g^b-g^h\,,\ (g^b+g^h)'\biggr)_{x=y=0.5}\,,
\end{split}
\eqlabel{mismatch}
\end{equation} 
with the superscripts $\ ^b$ and $\ ^h$ referring to the boundary (UV) and the horizon (IR) 
integrations, vanishes. At each iteration we adjust the  set of parameters \eqref{paruvir}
along the direction of the steepest decent for 
$||\vec{v}_{mismatch}||$. In practice, for a valid numerical solution we were able to  
achieve 
\begin{equation}
||\vec{v}_{mismatch}||\ \sim\  10^{-13}\cdots 10^{-11}\,.
\eqlabel{normachive}
\end{equation}

\subsection{Deformation of KT BH along $\c$SB tachyonic directions}

\begin{figure}[t]
\begin{center}
\psfrag{A}{{$A$}}
\psfrag{nv}{{$||\vec{v}_{mismatch}||$}}
\includegraphics[width=4in]{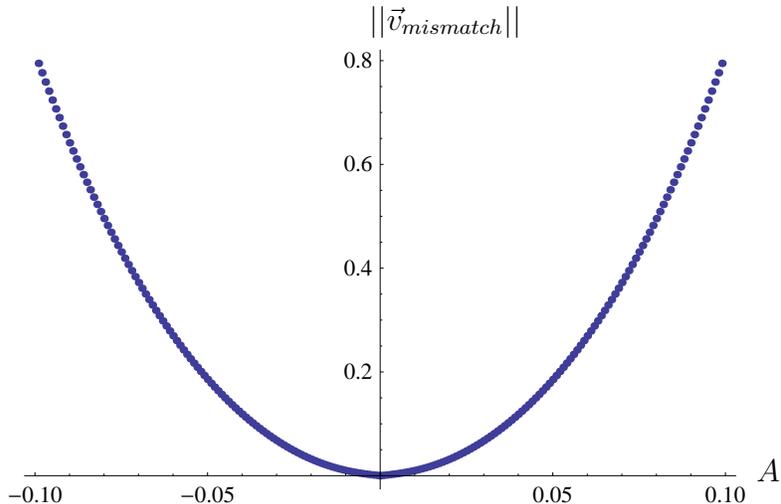}
\end{center}
  \caption{ A minimum of the mismatch vector $||\vec{v}_{mismatch}||$ \eqref{mismatch}
as a function of the 'tachyon deformation amplitude' $A$ \eqref{deform} 
for $A\ne 0$ would identify  seed values of  parameters \eqref{paruvir} 
leading to a homogeneous and isotropic KS BH solution with spontaneously broken chiral symmetry.
We use $k_s=-0.8$. Clearly, such minimum is not present.
} \label{figure3} 
\end{figure}

In section 3 we identified instabilities of the translationary invariant KT BH 
horizons, provided $T< T_{\c\rm{SB}}$. Earlier in this section we setup a general numerical boundary 
value problem to determine translationary invariant regular horizon geometries with 
 spontaneously broken chiral $U(1)$
symmetry. Here, we outline our attempts to construct such geometries.

First of all, since we are interested in {\it spontaneous} as opposite to {\it explicit} $\c$SB we set 
the mass deformation parameters (see \eqref{solfixed}) to zero:
\begin{equation}
f_{a10}=K_{110}=0\,.
\eqlabel{masszero2}
\end{equation} 
Second, motivated by the analysis of section 3, we consider values of $k_s$, such that the temperature of the 
KT BH is below the temperature of the chiral tachyonic instability $T_{\c\rm{SB}}$, but is still above the temperature 
of the hydrodynamic instability $T_u$. This translates into the range 
\begin{equation}
k_s^{unstable}=-1.230(3)\qquad <\ k_s\ <\ \qquad k_s^{\c\rm{SB}}-0.77743(2)\,.
\eqlabel{rangeks}
\end{equation}

The main difficulty associated with solving the boundary value problem 
(observed also in the analysis in \cite{kt3,hyd3}) is that the basin of attraction
of the parameters \eqref{paruvir} resulting in the convergent iterative process for 
the steepest decent for the norm of the mismatch vector \eqref{mismatch} is quite narrow; moreover, 
it becomes more and more narrow as $k_s$ (or equivalently the temperature) decreases.  
In other words, to obtain a solution one has to have a pretty good guess for the seed (initial)
values of \eqref{paruvir}. Clearly, having a 16-dimensional parameter space 
this is a daunting task! Of course, identical problem\footnote{The only difference being that 
the corresponding parameter space there is 10-dimensional.} exists for finding the KT BH 
solution. It is instructive to recall how this issue was 
circumvented in  \cite{kt3,hyd3}:
\nxt from general field theoretic arguments, \ie high-temperature restoration of the spontaneously 
broken symmetry,  
KT BH was supposed to exist at arbitrary high temperatures  \cite{b};
\nxt for $T\gg \Lambda$ one can develop an analytic high-temperature solution for the 
KT BH \cite{kbh2};
\nxt from the analytic high-temperature solution we can extract the values of the 
parameters for the boundary value problem and use them as 'seeds' \cite{kt3};
\nxt finally, we can slowly lower the temperature using as 'seeds' parameters 
obtains from solution of the boundary value problem at previous (slightly higher) 
temperature.\\
Rather remarkably, a described procedure, for small enough temperature decrements --- typically   
$\frac{\delta k_s}{k_s}\sim 10^{-2}$ ---  
resulted in convergence of the norm of the mismatch vector from initial values of 
order $10^{-1}$ to values \eqref{normachive} in 8 or less iterations.   

Given that the instability of the KT BH towards generating chiral condensates  
exists only below certain temperature, there is no high-temperature (analytic) guide
for the seed values of \eqref{paruvir}. In the rest of this section we explain one of our 
unsuccessful attempts to produce KS BH solution. As we emphasized before, in the limit of vanishing 
masses \eqref{masszero2}, for every values of $k_s$ we should recover the appropriate 
KT BH solution. Indeed, this is what we found: for instance, for $k_s=-0.8$ we 
recovered (with precision $\sim 10^{-8}$) KT BH parameters (we need to use \eqref{ktbh2}-\eqref{ktbh8}
and \eqref{ksircond})  with 
\begin{equation}
f_{a30}\ \sim\ k_{230}\ \sim\ f_{a70}\ \sim 10^{-9} \,,
\eqlabel{chiralvevex}
\end{equation}      
where the exact (expected) values should vanish \eqref{ktbh1}\footnote{This provides a highly 
nontrivial consistency check on our analysis.}.
Our idea was to start with a KT BH solution and deform its set of parameters
with the  parameters of the linearized $\c$SB tachyon  \eqref{normalizableuv} and 
\eqref{normalizableir} at amplitude $A$ (see \eqref{deffl}). At the level of 
functions \eqref{functions},
\begin{equation}
\begin{split}
&K_1=K^{KT}+A\ \dd k_1\,,\qquad K_2=1+A\ \dd k_2\,,\ K_3=K^{KT}-A\ \dd k_1\,,\\
&f_a=f_3^{KT}+A\ \dd f\,,\qquad f_b=f_3^{KT}-A\ \dd f\,,\\
&f_c=f_2^{KT}\,,\qquad h=h^{KT}\,,\qquad g=g^{KT} \,,
\end{split}
\eqlabel{deform}
\end{equation}
where the fluctuations $\{\dd f\,, \dd k_1\,, \dd k_2\}$ (computed 
at the threshold of instability) 
are substituted with $\kk=0$\footnote{We need to substitute $\kk=0$, otherwise the seed
 functions 
\eqref{deform} do not describe homogeneous and isotropic horizon. }. A linearized tachyon deformation
\eqref{deform} is off-shell now (since $\kk$ should vanish), but it is only slightly 
off-shell\footnote{If a homogeneous and isotropic KS BH 
horizon exists.} if $k_s$ is close enough to $k_s^{\c\rm{SB}}$. Physically, what we are doing is to allow 
the KT BH chiral tachyon to roll and build up the $\c$SB condensates of amplitude $\sim A$. 
The expectation is that as we scan the 'seeds' constructed from \eqref{deform} as a function of 
$A$ we should reach a new basin of attraction, different from the one of the 
KT BH solution,  in the parameter space \eqref{paruvir}. The iterative procedure in this new 
basin of attraction (as described at the end of section 4.4) would produce a KS BH solution.
A signature of a new basin of attraction would be a minimum of the mismatch 
vector constructed from the 'seed' parameters  from \eqref{deform} as a function of $A$, 
for $A\ne 0$.  Figure~\ref{figure3} presents typical results of such 
analysis\footnote{The norm of mismatch vector very rapidly and monotonically increases to $\sim 10^2$
as $|A|$ increases to $0.3$.}. We used 
$k_s=-0.8$ for data in Figure~\ref{figure3}.  The absence of a minimum in $||\vec{v}_{mismatch}||$
away from $A\ne 0$ is one piece of the evidence that 
a homogeneous and isotropic KS horizon with spontaneously broken chiral symmetry does not 
exist. Since $\c$SB fluctuations of the KT horizons are tachyonic for $T<T_{\c\rm{SB}}$, these 
tachyons must condense with finite momenta, resulting in non-homogeneous and non-isotropic ground 
state.  In section 5 we present independent analysis pointing to the same conclusion.

\section{Homogeneous and isotropic states of mass-deformed cascading plasma}

\begin{figure}[t]
\begin{center}
\psfrag{fa30}{{$f_{a30}$}}
\psfrag{fa10}{{$f_{a10}$}}
\psfrag{K110}{{$K_{110}$}}
\includegraphics[width=3in]{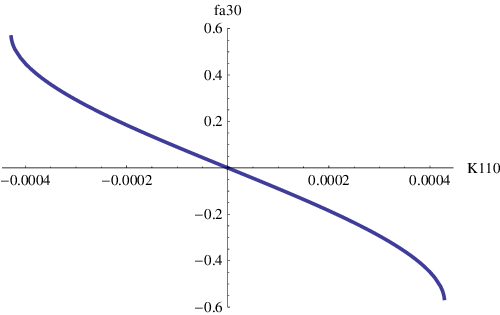}
\includegraphics[width=3in]{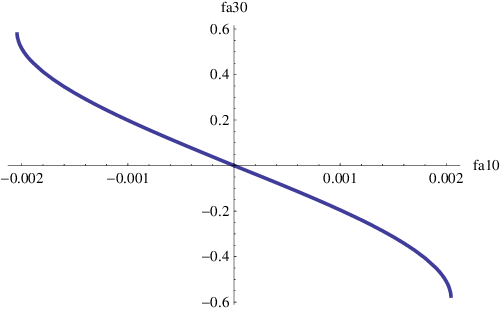}
\end{center}
  \caption{ One of the chiral condensates ($f_{a30}$) in mass-deformed cascading plasma as 
a function of mass-parameters $K_{110}$ with $f_{a10}=0$ (left plot) and $f_{a10}$ 
with $K_{110}=0$ (right plot) 
(see \eqref{solfixed} for the precise relation to gaugino masses) for $k_s=-0.8$. 
Notice that the condensate vanishing linearly in the chiral limit.
 } \label{figure4}
\end{figure}

We argued in section 3 that chirally symmetric deconfined phase of the cascading plasma 
becomes unstable with respect to fluctuations spontaneously breaking the chiral symmetry, 
provided $T<T_{\c\rm{SB}}$. We further presented the evidence in section 4.5 that these 
tachyons do not condense at zero momentum in a new ground state --- in other words, 
the metastable\footnote{Recall that $T_{\c\rm{SB}}$ is below the temperature of the 
first order confinement/deconfinement phase transition in cascading plasma.} equilibrium phase of  
deconfined cascading plasma at $T<T_{\c\rm{SB}}$ breaks the chiral $U(1)$ symmetry spontaneously, but is not 
homogeneous and isotropic.  In this section we present alternative arguments, leading to the same conclusion.

Effective gravitational action \eqref{5action} can describes homogeneous and isotropic 
thermal states of mass-deformed cascading plasma (see section 4.4). Specifically, we can introduce two 
independent mass-parameters $\mu_i=\frac{m_i}{\Lambda}\,, i=1,2$ 
related to the non-normalizable coefficients $\{f_{a10},K_{110}\}$ (see \eqref{solfixed}) of the 
general asymptotic UV expansion \eqref{uv1}-\eqref{uv8} of the holographically dual gravitational 
background. These are the mass terms for the gauginos of the cascading gauge theory $\caln=1$ 
vector multiplets. Gaugino mass terms explicitly break chiral symmetry, and thus the thermal 
state of the mass-deformed cascading gauge theory should exist at arbitrary high temperatures.
In particular, as recalled in section 4.5, we can now follow the strategy of constructing 
mass-deformed KS BH solution by developing first a mass-deformed high-temperature expansion, 
and then using obtained values of normalizable coefficients as 'seeds' for  \eqref{paruvir}.
Slowly varying $k_s$ we can reach low temperatures. Finally, we can numerically consider the limit 
of vanishing masses and study whether or not the condensates $\{f_{a30}\,, k_{230}\,, f_{a70}\}$ 
survive the chiral limit.    

We present only the final results\footnote{Much like in case of KT BH solution \cite{kt3}, consistency 
of the 'full solution' and its high-temperature limit for $T\gg \Lambda$ is a highly nontrivial check.}.   
We consider $k_s=-0.8$, which corresponds to temperatures below the condensation of the $\c$SB fluctuations,
see \eqref{rangeks}. Figure~\ref{figure4} presents results for the chiral condensate $f_{a30}$ 
(the remaining chiral condensates $\{k_{230}\,, f_{a70}\}$ have identical qualitative behavior)
in two special\footnote{We tried other mass-deformations, and the results are qualitatively identical.}
cases:
\begin{equation}
\begin{split}
&{\rm{left\ plot}}:\qquad (K_{110}\ne 0\,, f_{a10}=0)\,,\\
&{\rm{right\ plot}}:\qquad (K_{110}= 0\,, f_{a10}\ne 0)\,.\\
\end{split}
\eqlabel{cases}
\end{equation} 
The map between the gravitational parameters $\{f_{a10}\,, K_{110}\}$ and the mass-parameters 
of the deformed cascading plasma is given by \eqref{solfixed}.
Notice that in both cases, in the chiral limit, the condensates vanish linearly with the mass parameter:
\begin{equation}
f_{a30}\ \propto K_{110}\to 0\,,\qquad {\rm{or}}\qquad f_{a30}\ \propto f_{a10}\to 0\,.
\eqlabel{chirallim}
\end{equation}
Thus, we conclude that homogeneous and isotropic states of the deconfined cascading plasma do not 
break chiral symmetry spontaneously.

\section*{Acknowledgments}
After reading a draft of \cite{b} in 2000, Joe Polchinski commented that it 
would be nice to construct a Klebanov-Strassler black hole solution
to supplement a discussion presented there. I have been interested in this problem 
since then. Over the years I had valuable discussions on the subject with 
Micha Berkooz, Sunny Itzhaki, Shamit Kachru, Igor Klebanov, Leo Pando Zayas, 
Joe Polchinski, Rob Myers, Andrei Starinets,  Arkady Tseytlin and Amos Yarom.
I particularly benefited from the discussions and collaboration 
with Ofer Aharony. I would like to thank Aspen Center for Physics,
Galileo Galilei Institute, 
Kavli Institute for Theoretical Physics and  Weizmann Institute of Science 
and  for hospitality during the various stages 
of this project.
Research at Perimeter Institute is
supported by the Government of Canada through Industry Canada and by
the Province of Ontario through the Ministry of Research \&
Innovation. I gratefully acknowledge further support by an NSERC
Discovery grant and support through the Early Researcher Award
program by the Province of Ontario.

\end{document}